\documentclass[usenatbib,fleqn,letters]{mnras}

\usepackage{newtxtext,newtxmath}
\usepackage[T1]{fontenc}
\usepackage{ae,aecompl}
\usepackage[pdftex]{graphicx}
\usepackage{epstopdf}

\usepackage{amsmath}	
\usepackage{amssymb}	

\usepackage{ctable}
\usepackage{url}
\usepackage{xspace}
\usepackage[normalem]{ulem}

\def\msun{{\rm\,M_\odot}}

\newcommand{\kms}{\, {\rm km\, s}^{-1}}

\newcommand{\be}{\begin{equation}}
\newcommand{\ee}{\end{equation}}

\def\h2{${\rm\,H_2}$}

\newcommand\sref[1]{\hyperref[#1]{Section~\ref*{#1}}}
\newcommand\fref[1]{\hyperref[#1]{Fig.~\ref*{#1}}}
\newcommand\Eqref[1]{equation~(\hyperref[#1]{\ref*{#1}})}
\newcommand\tref[1]{\hyperref[#1]{Table~\ref*{#1}}}

\def\ensm{$\rm E_{NSM}$~}
\def\tcoalesc{$\rm t_{coalesce}$~}
\def\msun{$M_{\odot}$~}
\def\erg{$\rm erg$~}
\def\mpch{\rm Mpc/h}
\def\kpch{\rm kpc/h}

\newcommand{\vmax}{v_{\rm max}}

\newcommand{\vpeak}{v_{\rm peak}}
\newcommand{\Mpeak}{M_{\rm peak}}

\title[Simulating Neutron Star Mergers as r-process Sources in UFDs]{Simulating Neutron Star Mergers as r-process Sources in Ultra Faint Dwarf Galaxies}
\author[Safarzadeh, Scannapieco]{
	\parbox[t]{\textwidth}{
	Mohammadtaher Safarzadeh\thanks{E-mail: mts@asu.edu} \& Evan Scannapieco} \vspace*{6pt} \\
	School of Earth and Space Exploration, Arizona State University, Tempe, AZ 85287-1404, USA\\
}

\usepackage[all]{hypcap}

\pubyear{2017}

\begin{document}
\label{firstpage}
\pagerange{\pageref{firstpage}--\pageref{lastpage}}
\maketitle

\begin{abstract} 

To explain the high observed abundances of r-process elements in local ultra-faint dwarf (UFD) galaxies, we perform cosmological zoom simulations that include r-process production from neutron star mergers (NSMs). We model star-formation stochastically and  simulate two different halos with total masses $\approx 10^8 M_{\odot}$ at $z = 6$. We find that the final distribution of [Eu/H] vs. [Fe/H] is relatively insensitive to the energy by which the r-process material is ejected into the interstellar medium, but strongly sensitive to the environment in which the NSM event occurs. In one halo the NSM event takes place at the center of the stellar distribution, leading to high-levels of r-process enrichment such as seen in a local UFD, Reticulum II (Ret II). In a second halo, the NSM event takes place outside of the densest part of the galaxy, leading to a more extended r-process distribution. The subsequent star formation occurs in an interstellar medium with shallow levels of r-process enrichment which results in stars with low levels of [Eu/H] compared to Ret II stars even when the maximum possible r-process mass is assumed to be ejected. This suggests that the natal kicks of neutron stars may also play an important  role in determining the r-process abundances in UFD galaxies, a topic that warrants further theoretical investigation.

\end{abstract}
 
 \begin{keywords}
Stars: Neutron -- Galaxies: Dwarf-- Stars: abundances 
\end{keywords}

\section{Introduction}

The production of elements heavier than zinc requires events with high fluxes of neutrons. These particles are captured by lighter nuclei at a rate that is slow or rapid with respect to subsequent beta decays, leading to neutron-capture processes  labeled s or r, respectively  \citep{Burbidge:1957hf,Sneden:2003fq,Sneden:2008cf}. While the production of s-process elements for heavy elements with $A>90$ is thought to primarily occur in asymptotic giant branch stars \citep{Busso:1999ig}, r-process elements  could be formed in core-collapse supernovae (SNcc)  or in neutron star mergers (NSMs), with an ongoing debate regarding the dominance of one mechanism over the other \citep{Argast:2004hg}.  A key difference between these mechanisms is the number of events required to produce the abundances observed today. Coalescing NSMs are calculated to eject $10^{-3}-10^{-2} M_{\odot}$ of r-process matter \citep{Rosswog:1999wz,Rosswog:2000nj}, which is orders of magnitude larger than the $10^{-6}-10^{-5} M_{\odot}$ ejected by SNcc, but their rate is also significantly lower than core-collapse rates \citep{Cowan:1991ca,Woosley:1994ih,Kuroda:2008dw,Wanajo:2013io}.

One discriminant between these two models is the abundance of r-process elements found within local ultra-faint dwarf galaxies \citep[UFDs,][]{Brown:2012jo,Frebel:2012ja,Vargas:2013ei} discovered in deep wide-area sky surveys \citep{Koposov:2015cw,Koposov:2015ep,Bechtol:2015bd}. Recently \citet{Ji:2016ja} obtained high-resolution spectra of nine stars in the local UFD Reticulum II (Ret II), and  found that SNcc are unable to account for the observed high r-process element abundances in this galaxy. Instead, a rare event such as a NSM \citep{Tanvir:2013bt,Wallner:2015de} or a magnetorotationally-driven supernova (SN) \citep{Winteler:2012fv,Wehmeyer:2015kl,Nishimura:2015fv} is required to explain the observed high stellar abundances of Eu and Ba. 
  
 Other constraints on the production of r-process elements have been obtained through studies of the Milky Way (MW) halo. \citet{vandeVoort:2015jw} carried out a zoom simulation of a MW like halo at $z=0$ and concluded that NSM events can explain the observed [r-process/Fe] abundance ratios assuming $10^{-2} M_{\odot}$ r-process mass is ejected into the ISM in each NSM event.   \citet{Shen:2015gc} studied the sites of r-process production by post-processing ``Eris'' zoom simulations at $z=0$. They found that r-process elements can be incorporated into low metallicity stars at very early times, a result that is rather insensitive to modest variations in delay times, the delay distribution, and merger rates. 

In this study, we focus on simulating one NSM event in the star formation history of two UFD candidates at high redshifts and compare  these results with local observations. We simulate two different halos: one in which star formation begins at $z\approx13$ and another in which star formation begins at $z\approx8,$ which is consistent with the suggested  wide redshift range for reionization to quench the star formation in UFDs \citep{Brown:2014jn}.  We study the statistics of the NSM event in terms of three key parameters: the coalescence time of the two merging neutron stars, the energy associated with the event, and the mass of r-process material that is released into the ISM.  These results provide joint constraints that can be used to rule out or provide support for currently favored models for the sources producing r-process elements.

The paper is organized as follows:  In \S2 we present the setup of the zoom simulations. In \S3 we present our results on the r-process abundance of the gas and stellar content of the galaxies and in \S4 we discuss the implications and give conclusions. 

\section{Simulation Setup}

We use \textsc{ramses} \citep{Teyssier:2002fj}, a cosmological adaptive mesh refinement (AMR) code, which implements an unsplit second-order Godunov scheme for evolving the Euler equations. \textsc{ramses} variables are cell-centered and interpolated to the cell faces for flux calculations, which are then used by a Harten-Lax-van Leer-Contact Riemann solver \citep{Toro:1994gu}. The code is capable of  advecting any number of scalar quantities such as metallicity, and we add our own scalar quantity to track r-process enrichment. 

The Initial Conditions for our simulations were provided by MUlti-Scale Initial Condition generator \citep[{\sc music},][]{Hahn:2011gj}.  The adopted $\Lambda{\rm CDM}$ cosmology parameters are based on Planck 2013 \citep{Collaboration:2014dt}
$\Omega_{\rm m}=0.308$, $\Omega_{\rm b}=0.048$, $\Omega_{\Lambda}=0.693$,
$\sigma_8=0.823$ and $n=0.96$, where the Hubble constant is $H_0=100{h\,\rm km}
\,{\rm s}^{-1}\,{\rm Mpc}^{-1}$ with $h=0.678$. 
The simulation volume, halo mass, and stellar particle mass were all chosen as described below.   

The stellar mass and circular velocity of Ret II are measured to be $\approx5\times10^3 M_{\odot}$ and $\approx35$ km/s, respectively \citep{Roederer:2016es}.  The star formation histories of UFDs indicate that nearly 3/4 of the entire stellar mass content of such galaxies is formed by $z\approx10$ and  $\approx80$\% of the stellar mass content is already formed by $z\approx6$ \citep{Brown:2014jn}.   Thus we conducted an initial simulation of a volume was chosen to be  $2 {\rm Mpc/h}$,  which is over 10 times the non-linear length scale at  $z\approx6,$ and a resolution of $256^3.$

Within this volume, dark matter halos were found through the HOP algorithm \citep{Eisenstein:1998in}. 
The current halo mases of UFDs are uncertain and believed to be  within $10^8-10^9$\msun depending on the assumed dark matter density profile \citep{Simon:2007ee,Bovill:2009hg,BlandHawthorn:2015ke}. \citet{BlandHawthorn:2015ke} suggest that $M=10^7$\msun at $z=z_{\rm reion}$ is the minimum halo mass that could correspond to the local UFDs such that the star formation can accumulate enough stellar mass prior to reionization and assemble a gravitational potential well deep enough to survive a SN blast.  Thus  we selected two halos with masses $\approx 10^8$ \msun at $z = 6$ as our UFD candidate progenitors in order to capture the minimum variance in our results due to star formation histories.   

After selecting the candidate halos, we performed a zoom simulation on two of the candidates. The Lagrangian box sizes within the parent $8~(\mpch)^3$ that we re-simulated for two of the candidate halos are $246 \times 308 \times 253~ (\kpch)^3$ and $449 \times 644 \times 609~ (\kpch)^3$.The dark matter particles were refined to effective values $1024^3$ (64 times finer in mass). The dark matter particle mass in the zoom regions is 676 \msun.

 
The stellar mass particle is $m_*=\rho_{\rm th}\Delta x^3_{\rm min} N$ where $\Delta x_{\rm min}$ is the best resolution cell size achievable and N is drawn from a 
Poisson distribution
\begin{equation}
P(N) = \frac{\bar{N}}{N!} \exp({-\bar{N}}),
\end{equation}
where 
\begin{equation}
\bar{N}=\frac{\rho \Delta x^3} {\rho_{\rm th} \Delta x^3_{\rm min}} \epsilon_*,
\end{equation}
where the star formation efficiency $\epsilon_*$ was set to 0.01 \citep{Krumholz:2007ig} in our simulations.
Setting $L_{\rm max}=19$ together with $n_*=10 $ H/cm$^{3}$ as the threshold for the star formation in the cells
results in the stellar particle mass of $\approx50 M_{\odot},$ a value that allows us to resolve the stellar mass content of such systems, while still being massive enough to host the two supernovae needed to create a neutron star binary.  
$L_{\rm max}$ is the maximum refinement level that is allowed in the simulation. The adopted value was 19 which given 
the box of 2 $\mpch$ on each side corresponds to a resolution of 5.5 pc that is kept at all redshifts. 

We adopted a delayed cooling scheme for the SN feedback as discussed in \cite{Stinson:2006id} and \cite{Dubois:2015ex}, with a dissipation time scale of $5\times10^4$ years
to resolve the cooling radius of the SN Sedov phase given the spatial resolution in our simulations which is $\approx 5.5$ pc.  Cooling was modeled following \citet{Dopita:1996cz} for $T>10^4 K$. Below $T=10^4 K$ , down to 2.8 K,  we adopt metal-line cooling from CLOUDY \citep{Ferland:1998ic}.

Star formation was limited to sites with overdensities $\Delta >200$ to avoid the formation of stars in non-virialized structures. Following the \cite{Kroupa:2001ki} Initial Mass Function (IMF), a 100 solar mass stellar particle hosts one SNcc on average. To properly model the statistics of SNe, we followed Poisson statistics with the mean number of isolated SNe equal to $N_{\rm Isolated}=(1-P_{\rm NSM}) \times m_* /100 M_{\odot}$ 
and the mean number of pair SNe equal to $N_{\rm Pair}= P_{\rm NSM} \times m_*/200 M_{\odot}$. Here, $P_{\rm NSM}$ captures the probability that a newly formed massive star has a companion in the same mass range, such that they make a neutron star merger (NSM) in a coalescence timescale (\tcoalesc). 

A range of different values for $P_{\rm NSM}(\approx10^{-4}-10^{-5})$ has been studied in \citet{Argast:2004hg}. However, as the stellar mass content of Reticulum II is only $\approx5\times10^3 M_{\odot}$, we carry out our simulations such that one NSM event occurs in the star formation history of our galaxies which effectively means $P_{\rm NSM}=2\times10^{-2}$. The occurrence of only one such event is suggested by \citet{Ji:2016ja} given the observed statistics of such systems in the MW halo. We carry out simulations with three different coalescence time for \tcoalesc (=1, 3 and 10 Myr). As our result turned out to be not sensitive to coalescence time within this range, we only present our results for \tcoalesc (=1 and 10 Myr).

In this study, we assumed that only massive short lived stars contributed to the production of r-process elements and slow s-process channels were not modeled. Consequently we did not model elements such as Ba that have both r-process and s-process origin. Also, we did not model SN Ia because of their long delay times of the order of 200-500 Myr \citep{Raskin:2009du}. Given the stellar particle mass ($\approx 50 M_{\odot}$), 50\% of all the stellar particles were assumed to host one SNcc. The SNcc were assigned stochastically to each stellar particle. Therefore half of the stellar particles have 1 SNcc ejecting ($M_{\rm ejecta}=10 M_{\odot}$) with a kinetic energy of $E_{\rm SN}=10^{51} {\rm erg}$, 10 Myr after the star is formed. The metallicity yield for each SNcc is set to $\eta_{\rm SN}=0.1,$ meaning one solar mass of metals is ejected in each SNcc event and we assume 5.4\% of all the metals ejected is in the form of Fe which is consistent with the composition of IGM gas at $z\approx6$ (private communications with Frank Timmes).  
There is a large uncertainty for the yields at redshifts of reionization and we instead modeled our yields in agreement with high redshifts observations of gas composition instead of the local fits. Therefore effectively every stellar particle that hosts a SNcc ejects $M_{\rm Fe}=0.054 M_{\odot}$ into the ISM. 
Comparing to the iron yield from different stellar masses \citep{Woosley:1995jn} which is summarized with the following fitting formula \citep{Shen:2015gc} 
\be
M_{\rm Fe}=2.802\times 10^{-4} (\frac{m_*}{M_{\odot}})^{1.864} M_{\odot},
\ee
our yield is similar to a yield from a 16 $M_{\odot}$ star. It should be noted that we did not model Pop III stars or assign a different yield to them \citep[e.g.][]{Sarmento:2017du}.

Finally we explored a range of NSM models, varying the energy (\ensm), coalescence timescale ($\rm t_{coalesc}$) , and ejected mass of r-process material ($M_r$) associated with the merger. In particular, the parameter space covered in this study is 
\begin{itemize}

\item Three different values of the NSM energy, \ensm$=10^{50},$ $3\times10^{50},$ and $10^{51} {\rm erg}$ that captures the range of kinetic energies released in a NSM event as studied in  SPH simulations by \citet{Piran:2013da} for a range of neutron star masses ($1M_{\odot}-2 M_{\odot}$) with negligible spin parameter. This parameter impacts the spread of r-process elements in the host galaxy and determines the dispersion of enriched stars with r-process elements in the galaxy.

\item Two different coalescence timescales for neutron star mergers, \tcoalesc= 1 and 10 Myr. 
10 Myr after a star is born, a SNcc occurs, or two SNcc in the case of the NSM particle. The coalescence time refers to the time after this explosion of the 2 SNcc.
The lower limit is suggested by \citet{Belczynski:2002gi} for their theory regarding new short-lived NS-NS subpopulation, which are tight binaries with very short merger times. Moreover, binary systems with short initial separation in highly eccentric orbits can also lead to such short coalescence timescale \citep{Korobkin:2012cp}. While much longer timescales are possible \citep[e.g.][]{Dominik:2012cw}, we set the maximum at 10 Myrs for this study because larger timescales ($\gtrsim 30$ Myr) would lead the NSM event to occur a system with a much larger stellar mass than Ret II given the star formation history of the halos that we simulated.

\end{itemize}
 
There is rather large range of r-process mass ($M_r$) possible to be ejected in a NSM event. The minimum value, $M_r=10^{-4} M_{\odot},$~is set from results of SPH simulation of neutron star mergers by \citet{Oechslin:2002iu} which take into account general relativistic effects in a conformally flat approximation, and the maximum value ($\approx4\times10^{-2}M_{\odot}$) was set by two different SPH simulations \citep{Korobkin:2012cp,Piran:2013da} \footnote[1]{See Table 1 of \citet{Piran:2013da} for the range of r-process material that is ejected in a NS binary simulation given the masses of each individual neutron star. }. With semi-newtonian potentials for NS-NS and NS-BH mergers, \citet{Just:2015kr} studied the fate of NS mergers and found an ejecta mass of $\approx0.004-0.02 M_{\odot}$ and $0.035-0.08 M_{\odot}$ for NS-BH mergers which is consistent with fully relativistic simulations of NSMs  \citep{Wanajo:2014ia}. Utilizing relativistic hydrodynamic simulations, \citet{Goriely:2011fa} predict $M_r\approx10^{-3}-10^{-2} M_{\odot}$  making them the main source of elements with mass number $A>140$ for merger rates of $\rm 10^{-5} yr^{-1}$. Since $M_r$ has no significant impact on the dynamics of the  simulations, we studied this parameter in post-processing.

We stopped our simulations when the stellar content of the galaxy reaches $\approx10^{4} M_{\odot}$ and assumed that further star formation will be quenched by reionization \citep{Efstathiou:1992iw,Quinn:1996dp,Klypin:1999ej}. The proper treatment of reionization would require a larger box than used in this study (e.g. box size $\approx 100 \rm Mpc$) which would be computationally prohibitive given the needed spatial resolution. Given this mismatch, it is impossible to carry out simulations of sufficiently high resolution to track single NS-NS mergers, and also self-consistently select galaxies whose overall star-formation terminates at the desired mass at the moment the wave of reionization overtakes their local environments.  For this reason, simply terminating the star formation is taken as the best possible approximation to the full process. 

Moreover, as we will show in (Safarzadeh \& Ji, in prep), halos with mass $\rm log M \approx 7.5$\msun at $z\approx8$ have $\approx10\%$ probability to survive as UFDs in MW progenitors. This result is based on merger tree analysis carried on the Caterpillar suite of zoom-in simulations \citep{Griffen:2016kn} on MW type halos and is consistent with the results presented previously in \citet{Gnedin:2006gl}. Therefore the halos selected in this study could be considered as having properties typical of the $\approx10\%$ of faint high-redshift dwarfs that survive to be observed as present day UFDs.

\section{Results}

Our fiducial simulation parameters are $E_{\rm NSM}=1\times10^{50}$ \erg, \tcoalesc= 10 Myr, and $M_r=10^{-3}  M_{\odot}.$  We simulate two separate halos: one in which  the star formation starts at $z\approx8$ and the other in which star formation starts at $z\approx13$. In both cases the stellar mass of the galaxy reaches that of Ret II after $\approx30$~Myr.

We convert the total r-process mass ejected into the ISM by a NSM event to Europium (Eu) abundance as 97.8\% of all Eu is produced by the r-process \citep[see table 5 of ][]{Burris:2000iz}. Fully relativistic simulations of NS mergers by \citep{Korobkin:2012cp} have shown that the electron fraction ($Y_e$) of the ejecta  can have a rather large ($\approx2 $) dex impact on the abundance of an element like Europium. However, the abundance of r-process elements is robust to the changes in properties of the NSM event such as the binary mass ratio in a NS-NS merger with a low $Y_e$ ($\approx0.04$). The solar abundance of Eu is A(Eu) = 0.52 \citep{Asplund:2009eu}  in the notation of $A(X)=\log (N(X)/N(H))+12$ and the mass fraction of Eu in total r-process mass ejected in a NSM event is \citep{Burris:2000iz,Argast:2004hg}:
\be
f_{\rm Eu}=\frac{N^r_{\rm Eu}.<m_{\rm Eu}>}{\Sigma_i N^r_i . <m_i>}\approx 1.2 \times 10^{-2},
\ee
where $N_i$ is the number fraction of r-process nuclei of the nuclear species $i$, and $<m_i>$ corresponds to
mean atomic weight. The sum goes over all elements beyond Ba.
In this work we have assumed all the Eu is generated in a single NSM event and the contribution from core-collapse supernovae is considered negligible. \citet{Matteucci:2014jt} 
have shown that neutron star mergers alone can explain both the gradient of the [Eu/H] along the galactic center and the production of Eu in MW. They suggest the coalescence timescale to be 
not longer than 1 Myr and the Eu yield per NSM event to be $\approx 2\times 10^{-7} M_{\odot}$. However, when they include the possibility of SNcc contribution to Eu production, coalescence timescales between 10-100 Myr are possible with similar Eu yield.  In our simulations, given the fiducial total r-process mass that is ejected in a single NSM event to be $M_r=10^{-3} M_{\odot}$, we are effectively assuming a Eu yield of $1.2 \times 10^{-5} M_{\odot}$ which is about 2 orders of magnitude larger than \citet{Matteucci:2014jt} suggest. However, assuming two orders of magnitude less Eu would result in too low [Eu/H] for our stellar particles (even for the system at $z\approx7$) to match the RetII system. 

Figure 1 shows the distribution of gas-phase, metals, and r-process elements in a simulated galaxy at $z=7.13$ with $\approx$ 360 stellar particles. Here the six panels illustrate the effect of varying the \ensm and \tcoalesc of the neutron star merger. Each row shows the result of increasing the \ensm from $10^{50}$ to $10^{51}$ \erg~ for a specific value of \tcoalesc. The red contours show the [Eu/H] and the black contours show [Fe/H] of the gas. The NSM is a stellar particle with a specific ID number that is the same for all simulations, however the assignment of SNcc to stellar particles is done in a stochastic fashion and therefore the contours of [Fe/H] can be different for each simulation. The colorbar shows the projected gas density in units of $\rm gr/cm^3$. This figure should be compared to our Figure 4 which we come into shortly.


\begin{figure*}
\centering
\resizebox{2.1in}{!}{\includegraphics[]{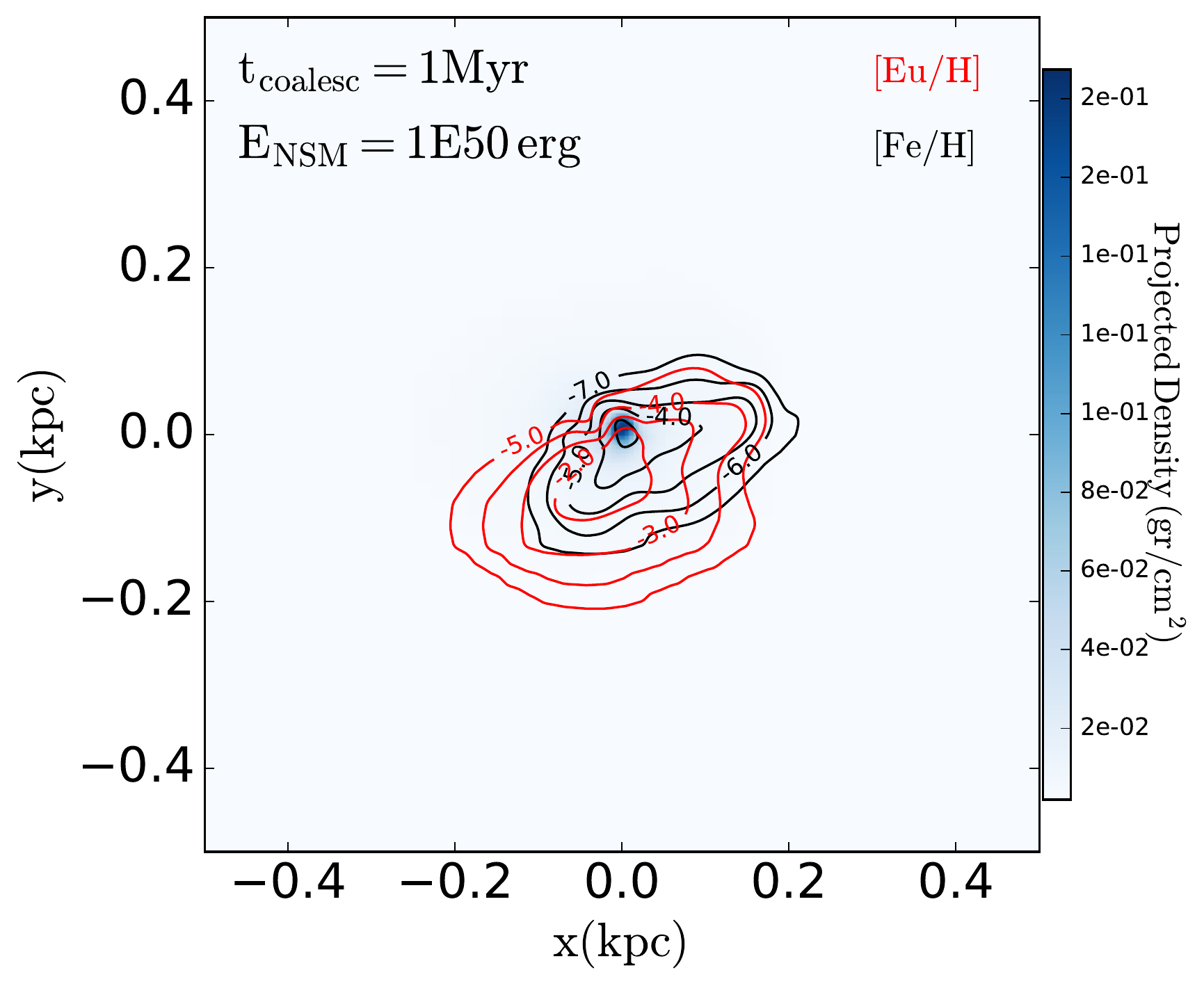}}
\resizebox{2.1in}{!}{\includegraphics[]{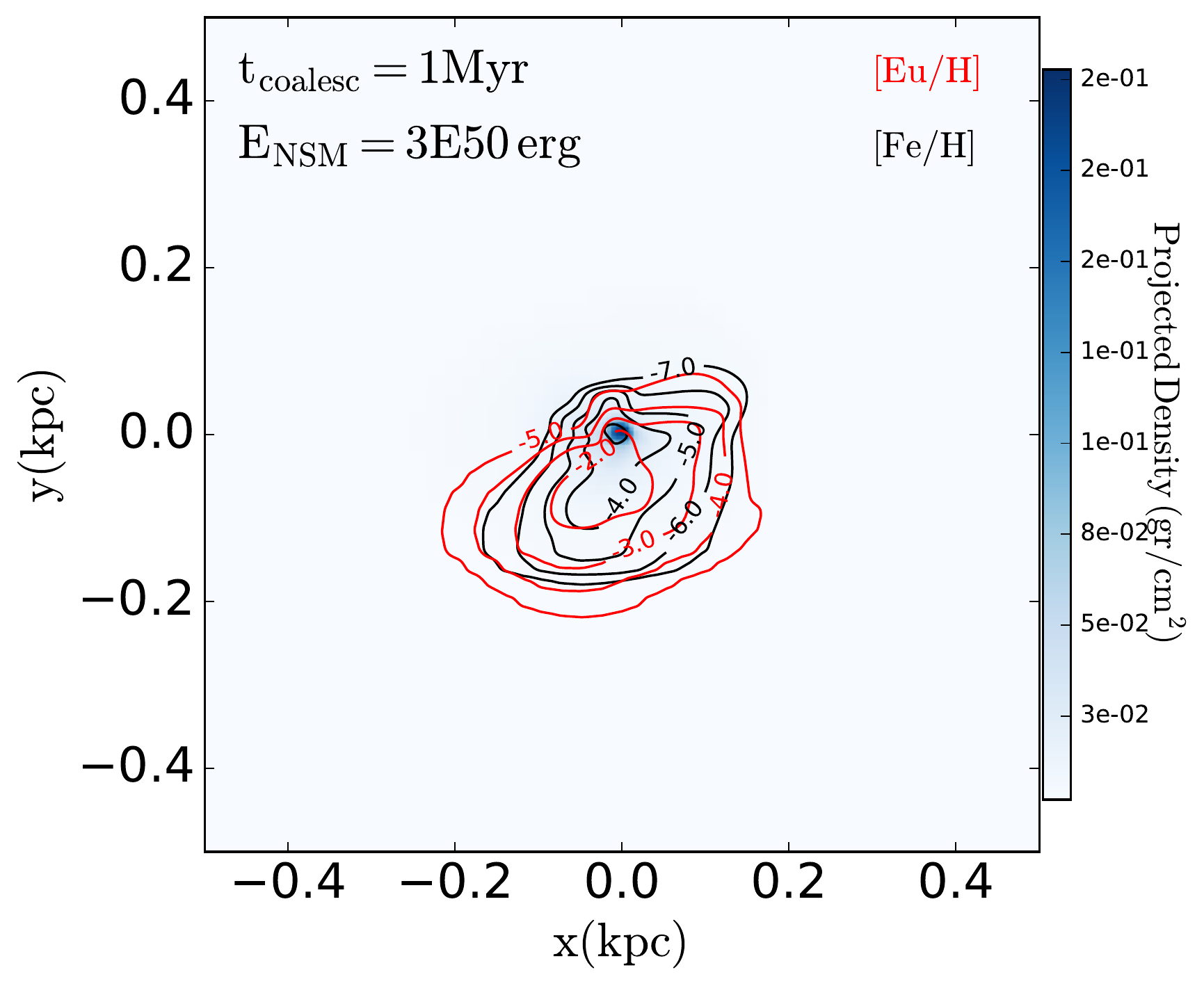}}
\resizebox{2.1in}{!}{\includegraphics[]{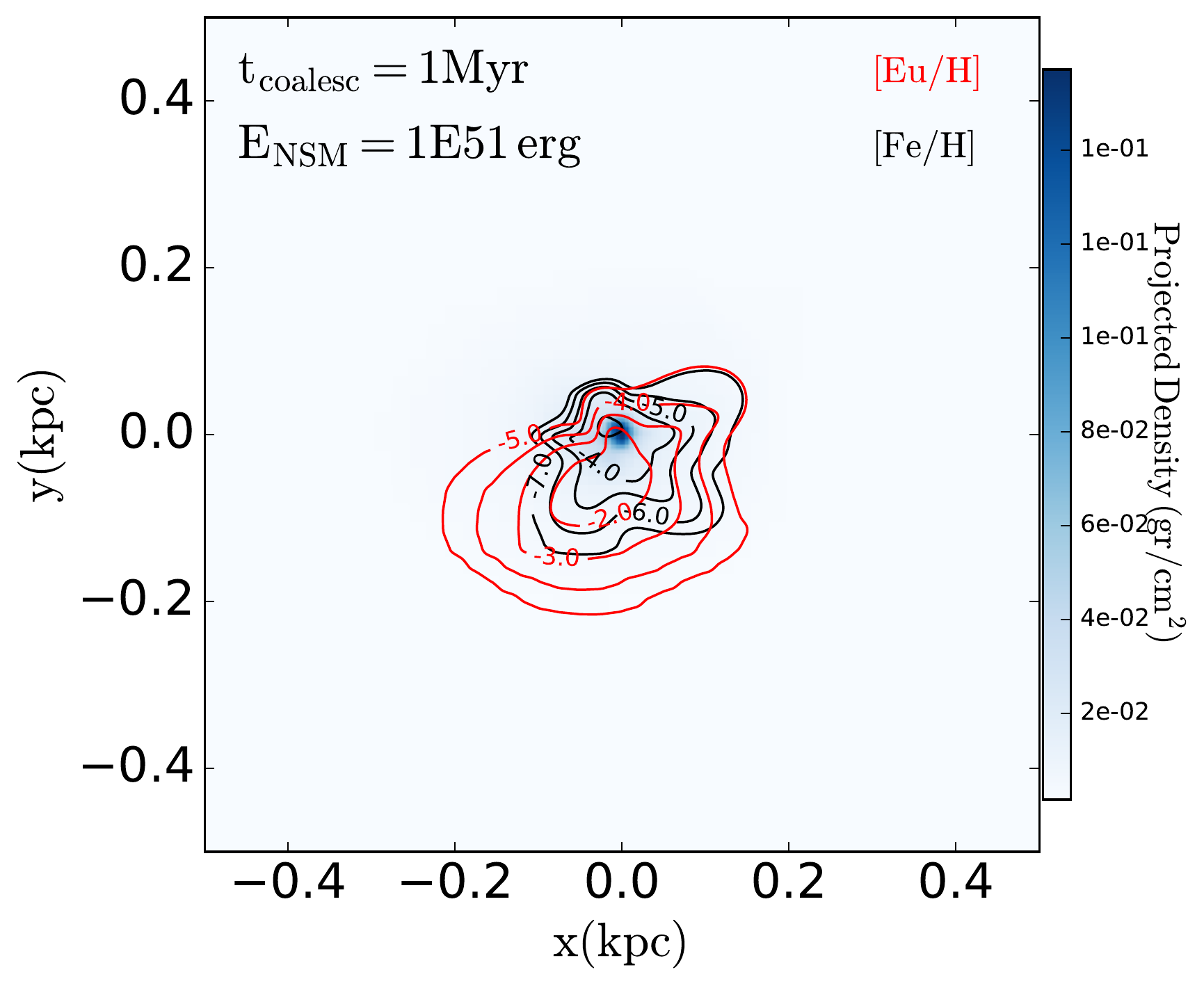}}
\resizebox{2.1in}{!}{\includegraphics[]{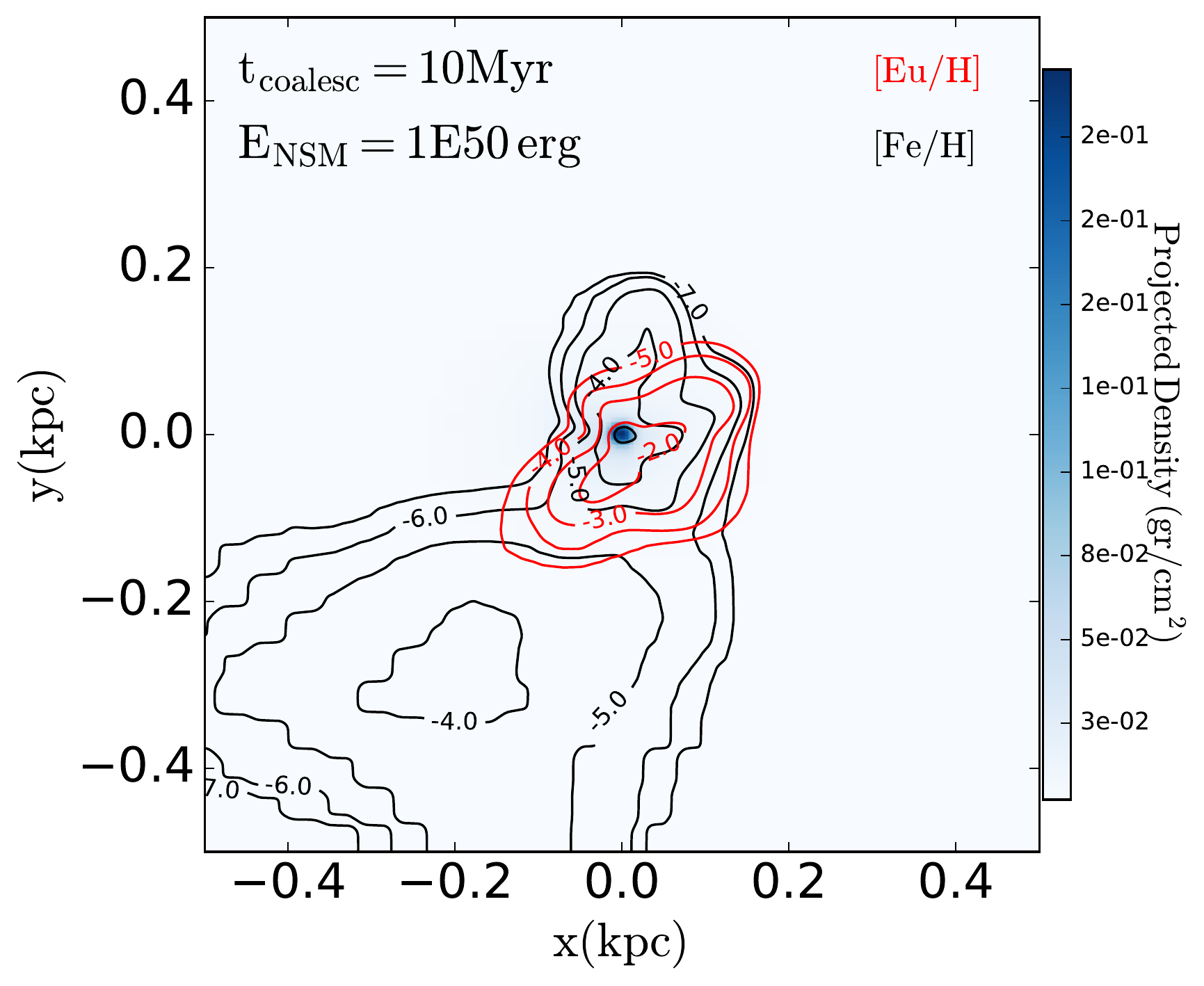}}
\resizebox{2.1in}{!}{\includegraphics[]{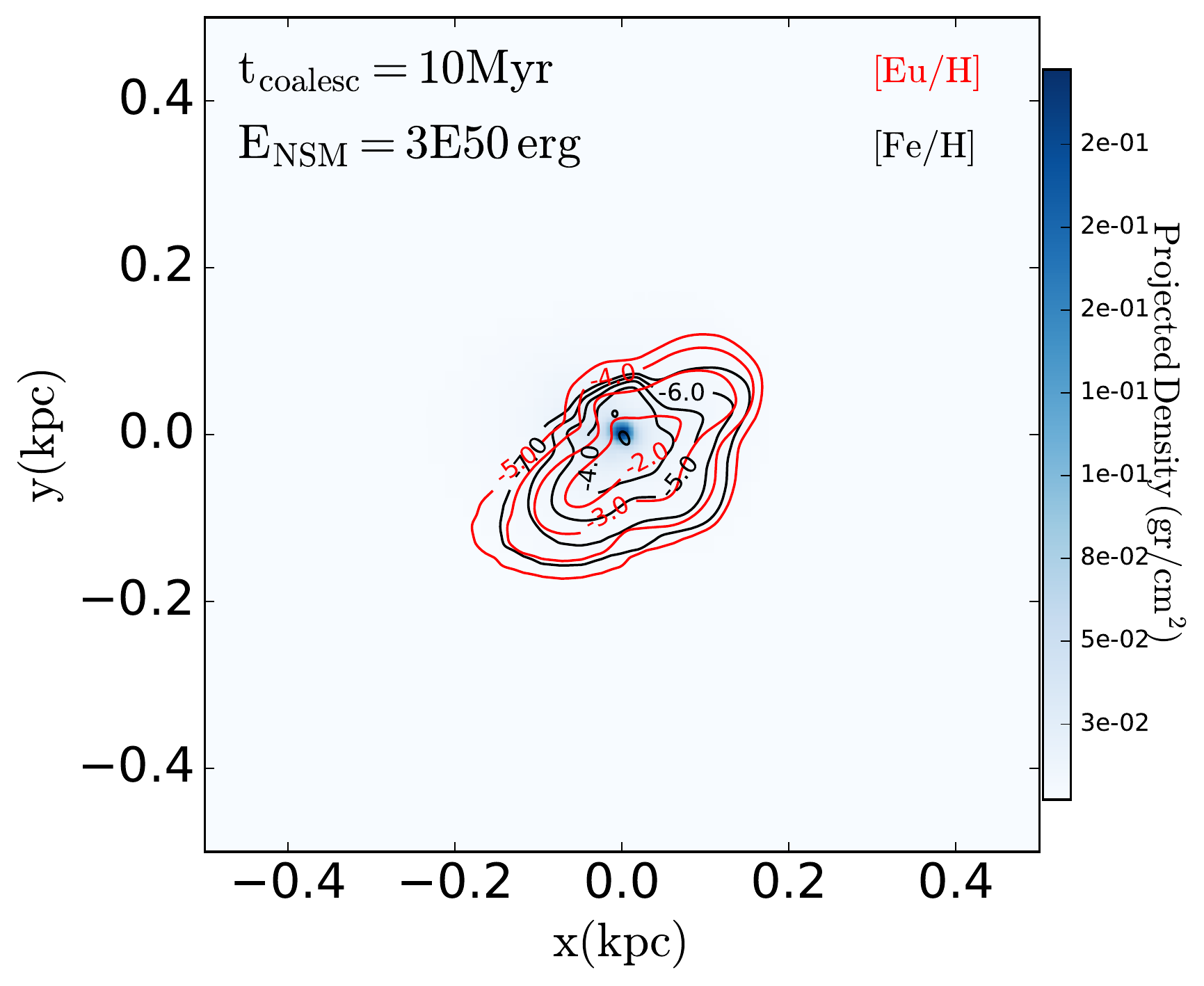}}
\resizebox{2.1in}{!}{\includegraphics[]{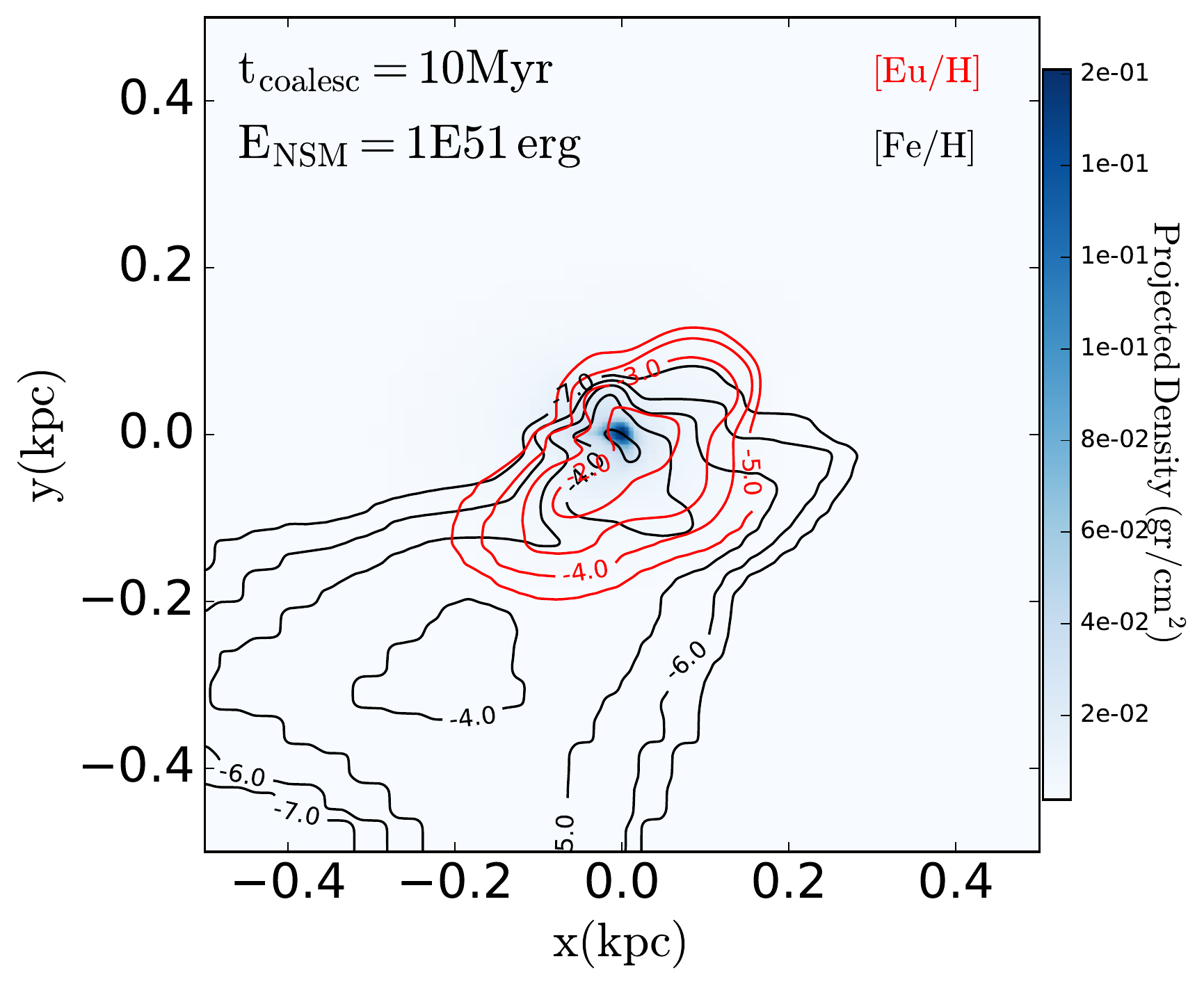}}
\caption{The projected density of our UFD candidate at $z=7.13$. Overlaid are contours of the gas phase abundances of [Fe/H] (black) and   [Eu/H] (red). 
From left to right \ensm = $10^{50}$, $3 \times 10^{50},$ and  $10^{51}$\,erg and from top to bottom \tcoalesc = 1 and 10\,Myr. 
In all cases the r-process material mass ejected in a NSM is set to $M_r=10^{-3} M_{\odot}$. 
When \tcoalesc =1 Myr, the NSM occurs at $z=7.4$,  therefore when  \tcoalesc=10 Myr, the NSM occurs 9 Myr after z=7.4. The NSM is a specific particle 
that is tagged in our simulation and therefore unique for a given set of simulation, however the assignment of SNcc to stellar particles is done in a stochastic fashion
and therefore the contours of [Fe/H] can be different for each simulation. }
\vspace{-0.3cm}
\label{low_z_gas}
\end{figure*}
 
Figure 2 shows the stellar particles in [Eu/H]-[Fe/H]  plane. The red points are the nine brightest stars in Ret II \citep{Ji:2016ja} which consist of seven detections and two upper limits.  The two most metal poor stars are upper limits on [Eu/H] (indicated as downward triangles), which might be indicative of multiple star formation epochs \citep{Webster:2015kb}. 
In all cases  [Eu/H] and [Fe/H] are positively correlated, which is consistent with the observations \citep{Burris:2000iz}.

The positive correlation between [Eu/H] and [Fe/H] is due the fact that  the NSM event and SNcc events are spatially correlated, in that both types of events happen in the dense part of the galaxy.  As stars with high [Eu/H] are those that form near the NSM, this means that they are also formed in regions that are highly-enriched with SNcc material.  Note that this trend can only be captured by modeling the spatial distribution of the ejecta.  In fact, we expect no correlation between  [Eu/H] and [Fe/H] if
the spatial location of the NSM event is not related to the SNcc events, even if these occur at the same time.  
We show this point in more detail below, presenting the results of the second halo we have analyzed.

\begin{figure*}
\centering
\resizebox{2.1in}{!}{\includegraphics[]{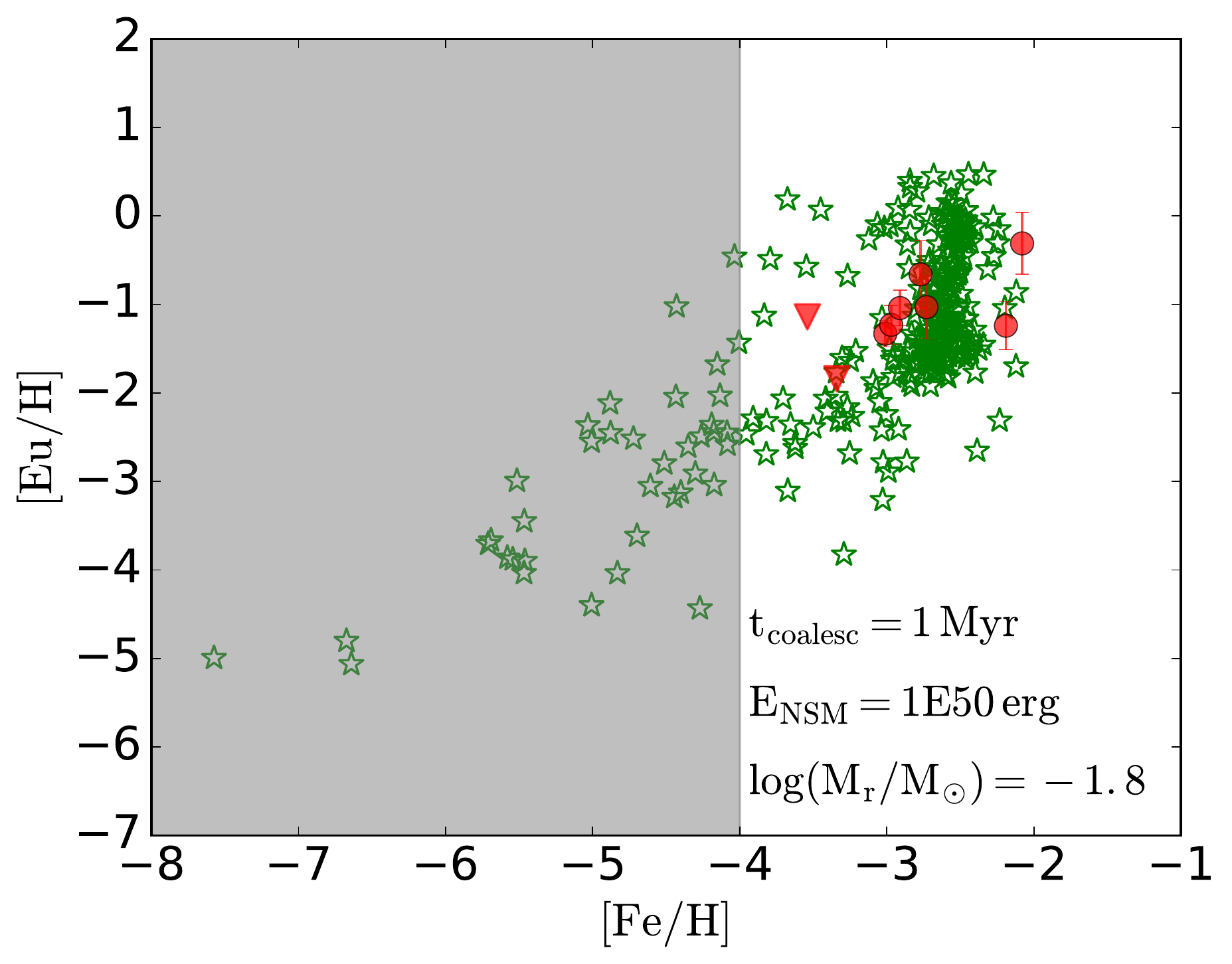}}
\resizebox{2.1in}{!}{\includegraphics[]{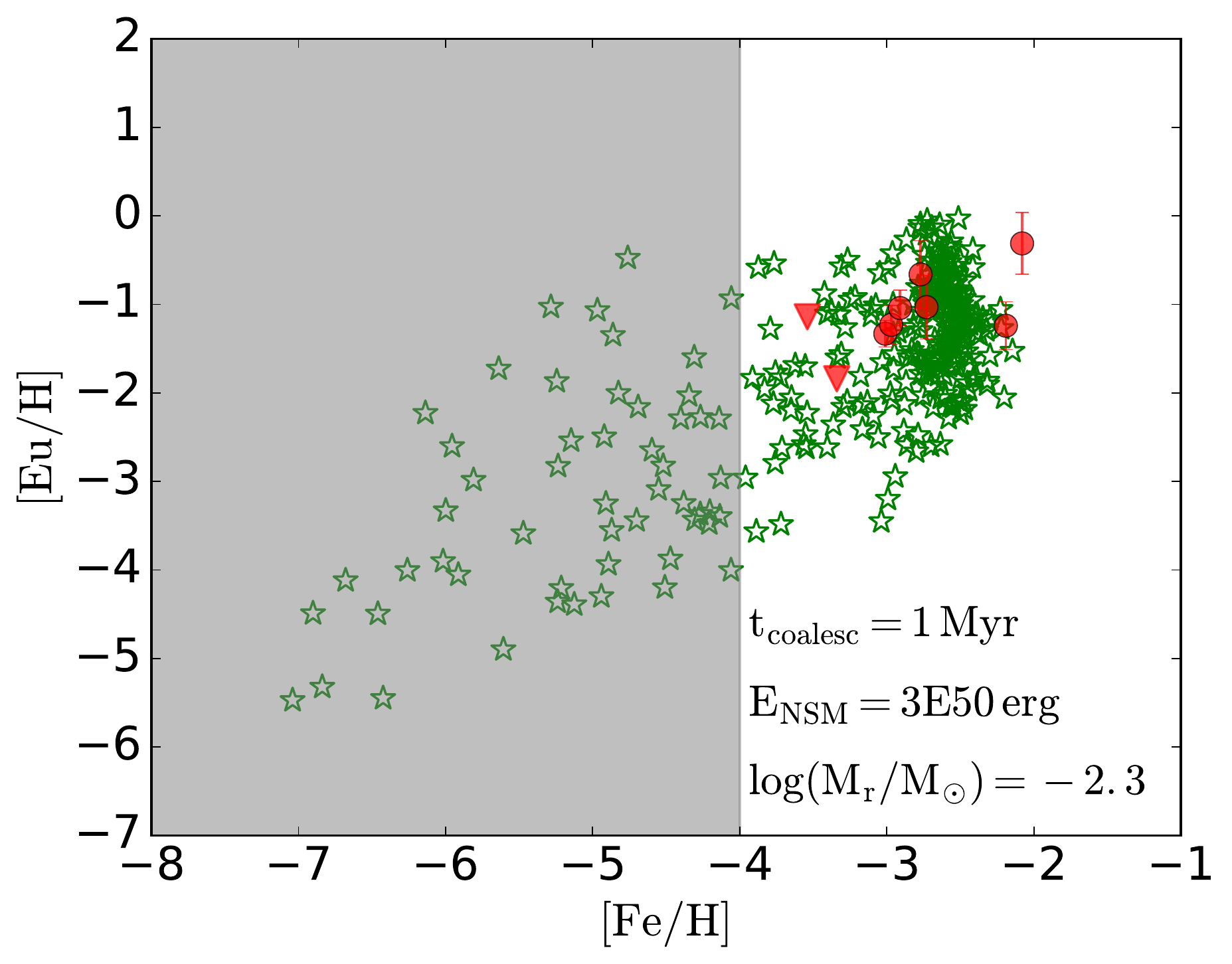}}
\resizebox{2.1in}{!}{\includegraphics[]{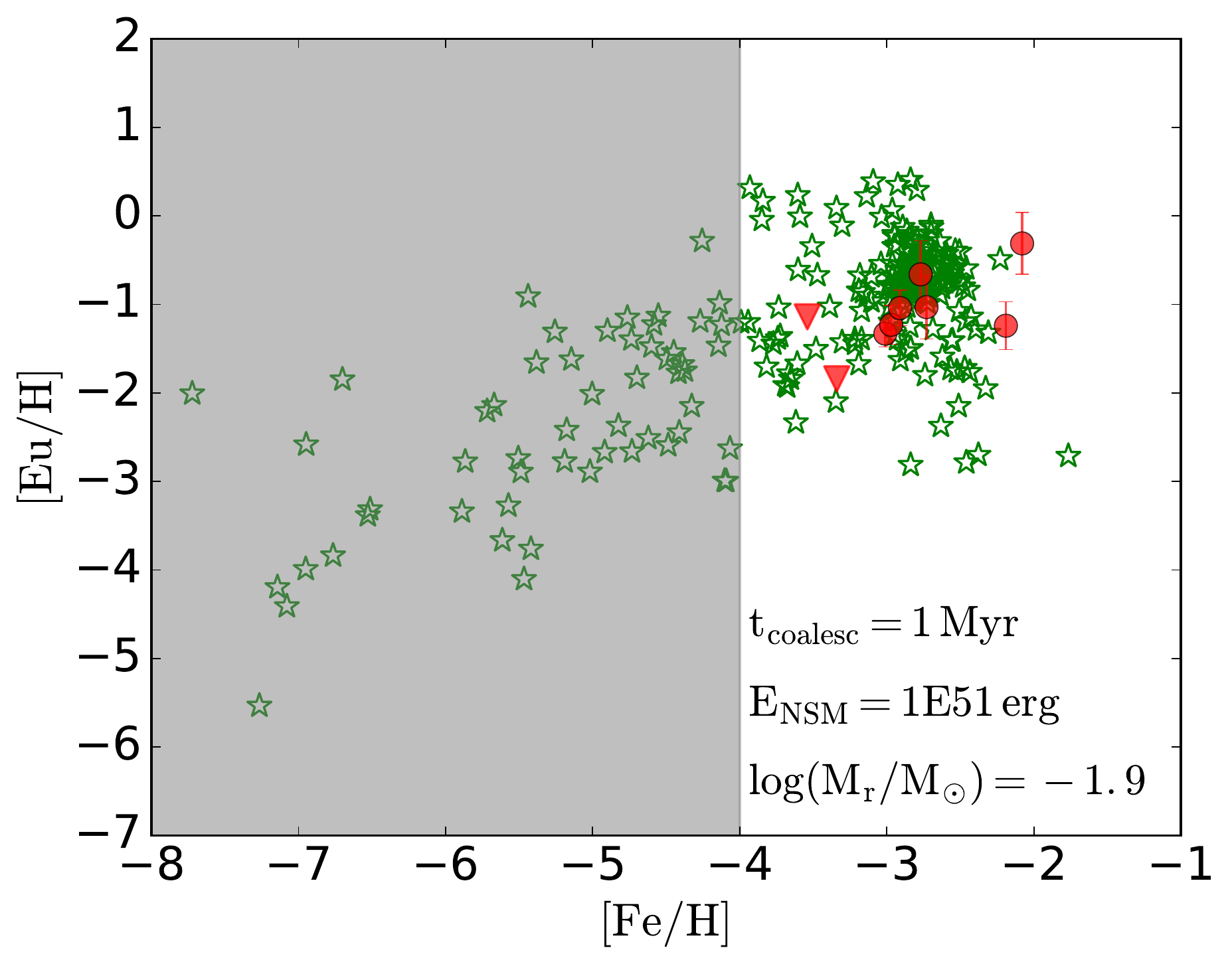}}
\resizebox{2.1in}{!}{\includegraphics[]{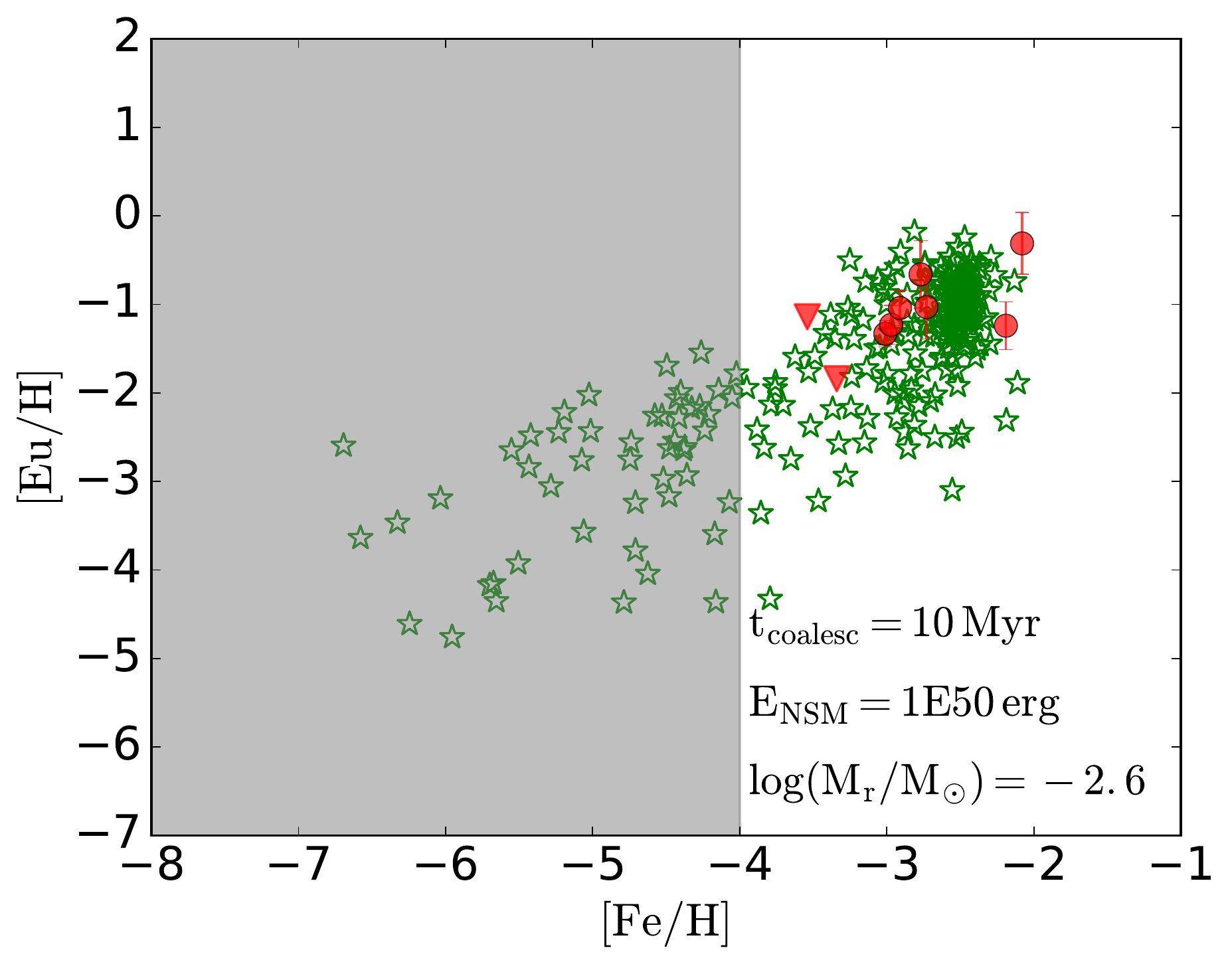}}
\resizebox{2.1in}{!}{\includegraphics[]{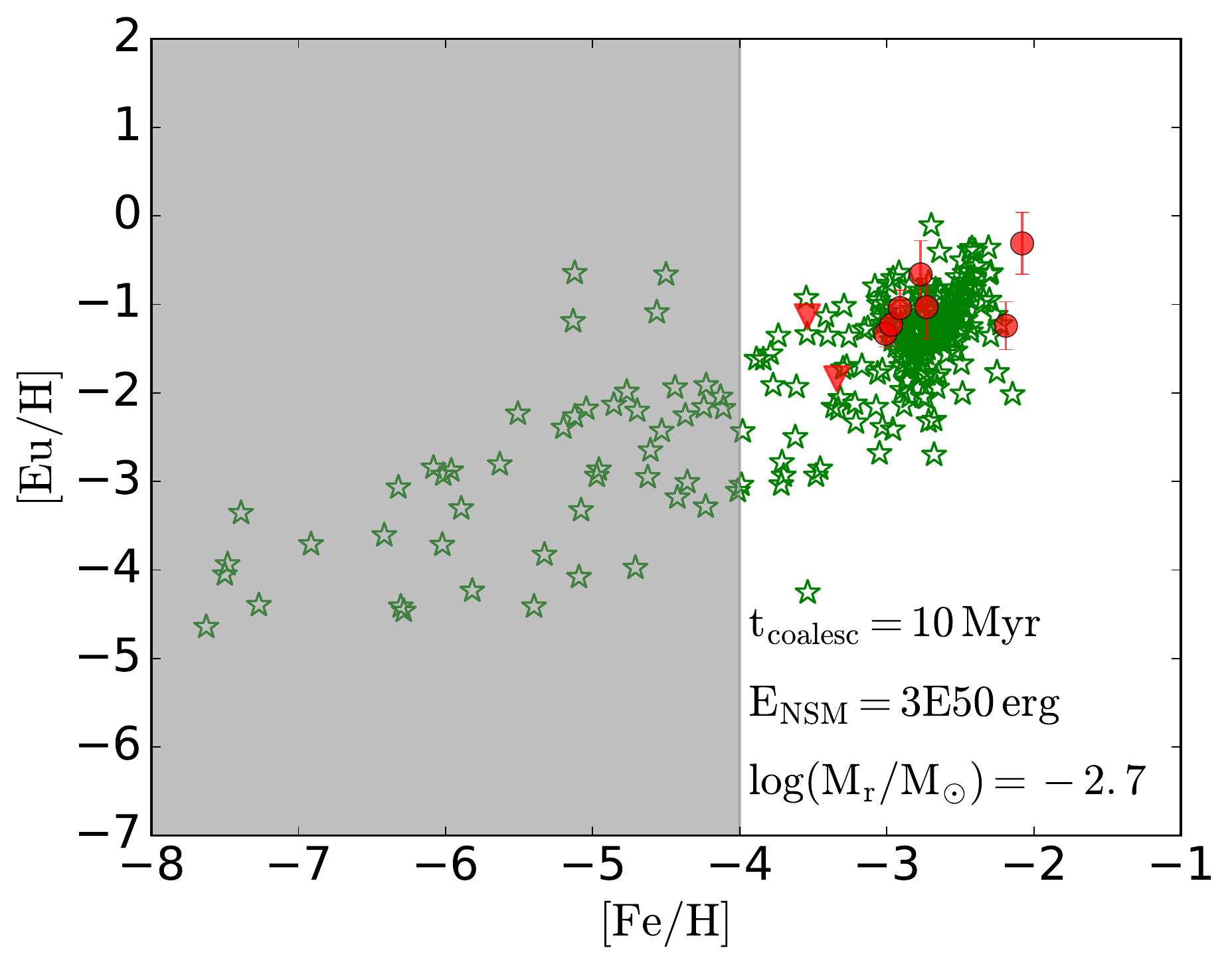}}
\resizebox{2.1in}{!}{\includegraphics[]{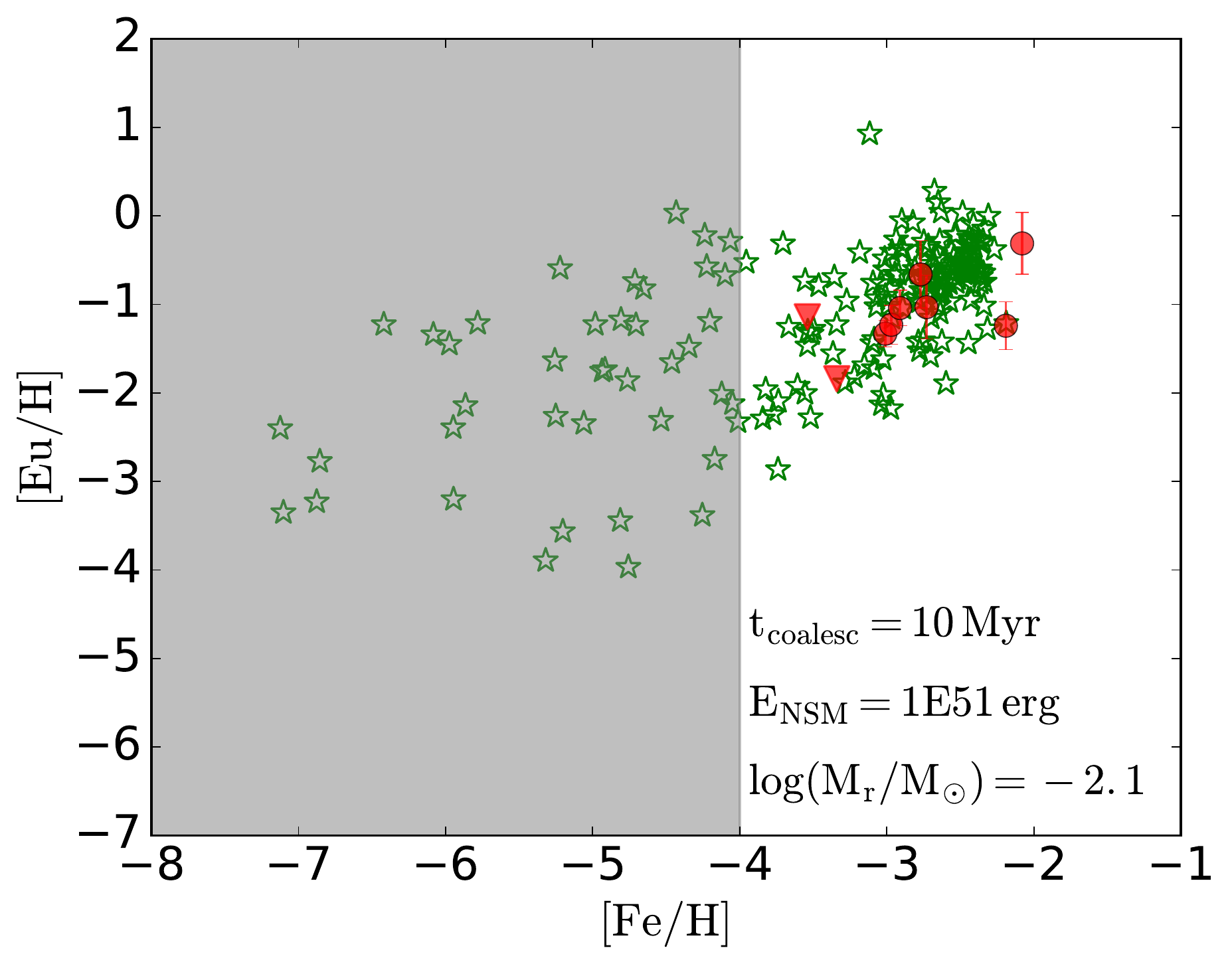}}
\caption{The [Eu/H] versus [Fe/H] for the stellar particles for our system at $z=7.13$ (open green stars) vs
 the seven stars heavily enriched by r-process  elements observed in Ret II  \citep[solid red points,][]{Ji:2016ja}.
 The two most metal-poor stars in Ret II are upper limits for [Eu/H], not detections.
As in Figure 1, from left to right \ensm = $10^{50}$, $3 \times 10^{50},$ and  $10^{51}$\,erg and from top to bottom \tcoalesc = 1 and 10\,Myr. In all cases, the r-process material mass ejected ($M_r$) in a NSM is mentioned in the panel.
The $M_r$ values are the ones that give the best 2D KS test P value when comparing the predictions and the observations in the 2D plane. The P values are all $\approx0.1$ and therefore we can not reject or prefer a model over the rest. The shaded region indicates stars that would belong to Pop III and therefore not be detectable today. The stellar particles in this region are not used to calculate the 2D KS test. }
\vspace{-0.3cm}
\label{low_z_stars}
\end{figure*}

In all simulations, we have adopted a fiducial value for the injected total r-process mass in a NSM event of $10^{-3} M_{\odot}$. However, the suggested range based on GR simulations of neutron star mergers spans the range between $10^{-4}-4\times10^{-2} M_{\odot}$. We therefore move the stellar particles in [Eu/H] accordingly to find the best P value that could be achieved by performing a 2D KS test against the seven detections (not including the two stars with upper limits on [Eu/H]). The stellar particles' Eu abundance is plotted for the value of $M_r$ that results in the best P value. 
In all cases we get a P value of $\approx 0.1$ and therefore can not reject any of our models or prefer one over the rest. However, we see that higher values of $M_r$ in the plausible range of $10^{-4}-4\times10^{-2} M_{\odot}$ are preferred to match the observations. We only consider stars with [Fe/H]>-4  for the KS test.

Our choice of parameters for the box size and stellar mass are chosen to resolve the Ret II dark matter halo and stellar content. At our fiducial resolution, the stellar particles have 
 a mass of about 50 $M_{\odot}$. Going one level lower in resolution would result in stellar particles larger in mass by a factor of 8, which would prevent us from resolving the 
 stellar content of the Ret II. Therefore to do the resolution study, we carried out the same simulation but this time with $L_{\rm max}=18$ compared to 19. This corresponds to a resolution of $\approx11$ pc compared to 5.5 pc. Since the stellar particle mass becomes 8 times more massive ($\sim400 M_{\odot}$), we decrease the threshold density for the star formation from $n_*=10 ~ \rm H/cm^3$ to $\rm 1.25~  H/cm^3$ to have the same mass for our stellar particles as our fiducial runs. This lower value for the threshold of star formation naturally leads to the formation of stars at earlier times. Figure 3 compares the results of our fiducial simulation (left panel) and the low resolution simulation (right panel) at $z=7.26$. As is shown, the results of r-process enhancement are little changed. Stellar particles are enriched to the same level of enhancement with $\log(M_r/M_{\odot})=-1.9$. Figure 3 shows the stellar particle r-process enhancement when we model the case with \ensm=$10^{51}$ erg and \tcoalesc=1 Myr with a lower resolution. The result for the lower resolution study is for the galaxy at $z=7.26$ with 155 stellar particles formed up to that redshift.
 
 We note that the turbulent cascades that lead to mixing in nature \citep{Pan:2010db} may not be sufficiently well captured if the AMR hierarchy is truncated at too low a maximum refinement level, possibly leading to over mixing of enriched material between cells \citep{Iapichino:2008fi,Iapichino:2008hw}.
 
\begin{figure*}
\centering
\resizebox{3.2in}{!}{\includegraphics[]{delayed_cooling_Eu_ID_5400z_7_13_1_1E51.pdf}}
\resizebox{3.2in}{!}{\includegraphics[]{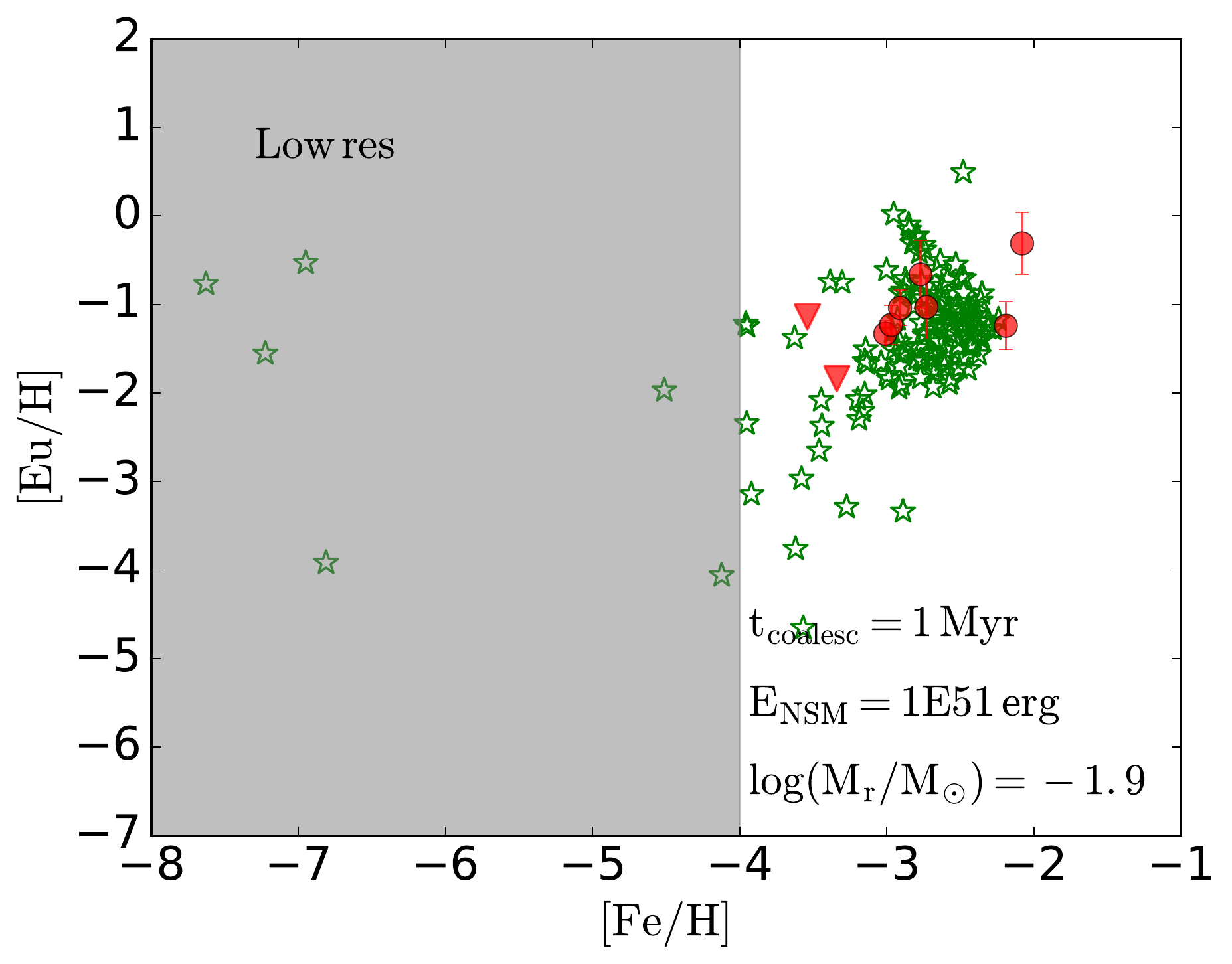}}
\caption{Showing the result of re-simulating one the case with \ensm=$10^{51}$ erg and \tcoalesc=1 Myr with our fiducial resolution on the left and with one level lower resolution corresponding to $L_{\rm max}=18$ on the right. Stellar particles are enriched to the same level of enhancement as for the higher resolution simulation with $\log(M_r/M_{\odot})=-1.9$. }
\vspace{-0.3cm}
\label{low_res_figure}
\end{figure*}

\begin{figure*}
\centering
\resizebox{2.1in}{!}{\includegraphics[]{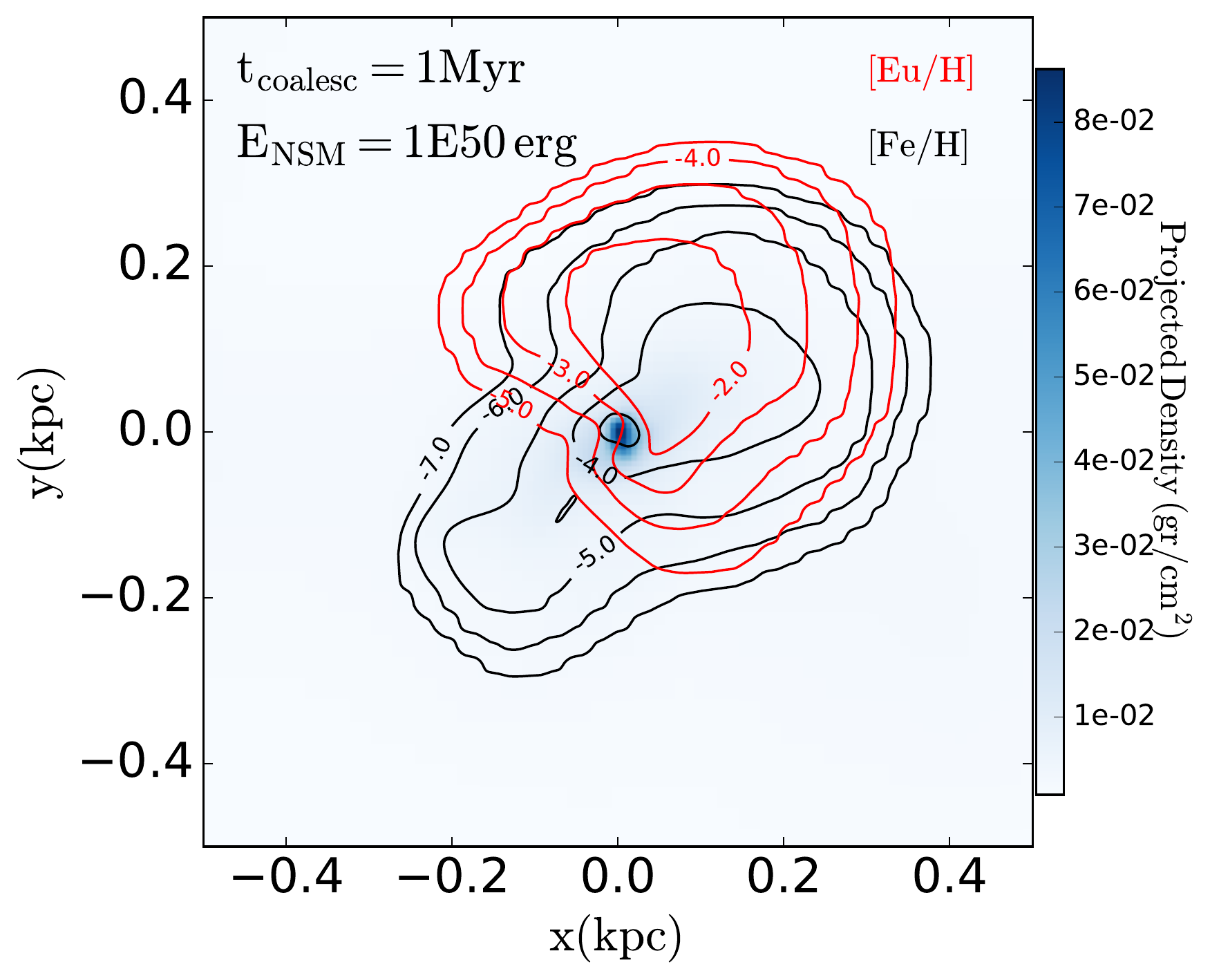}}
\resizebox{2.1in}{!}{\includegraphics[]{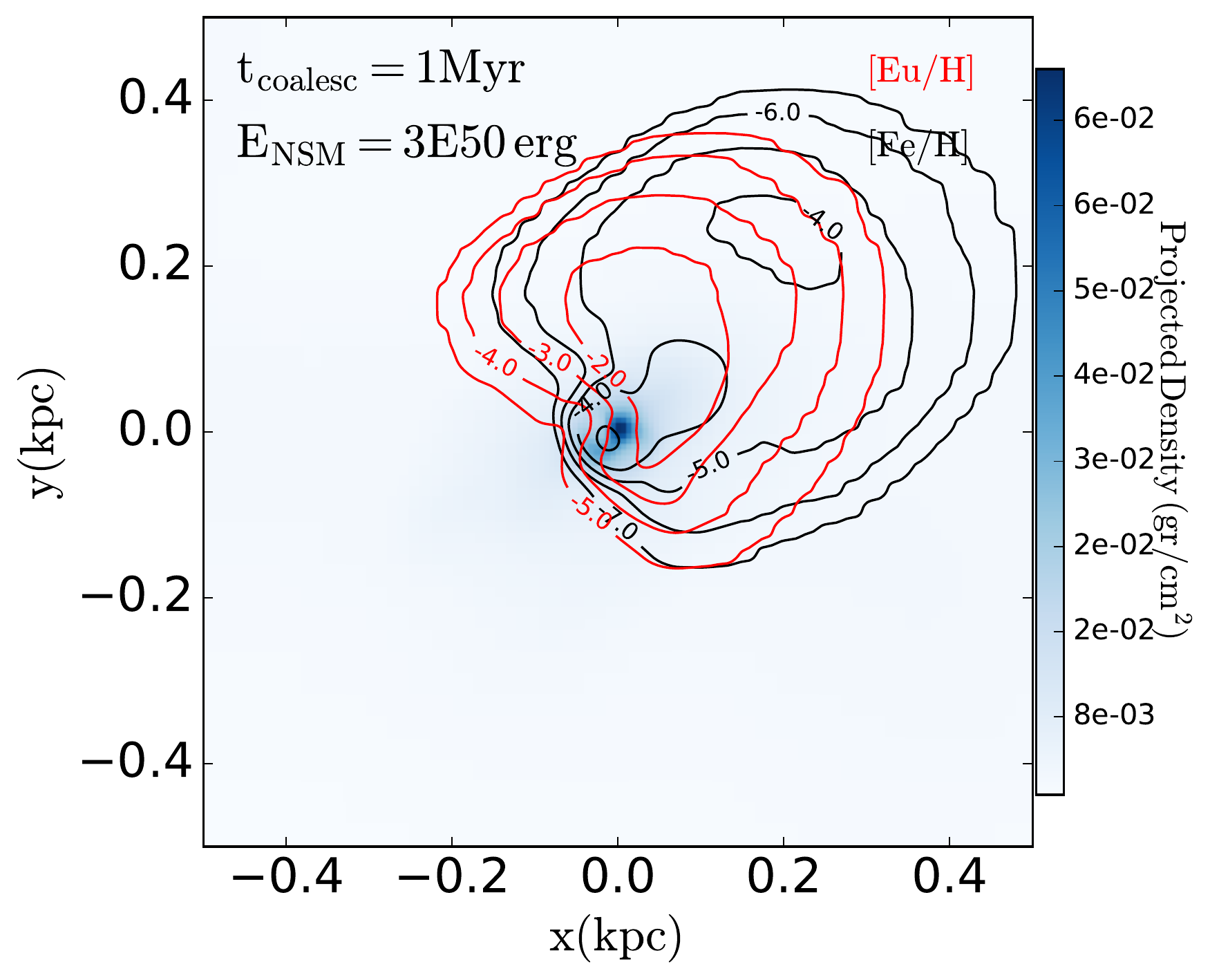}}
\resizebox{2.1in}{!}{\includegraphics[]{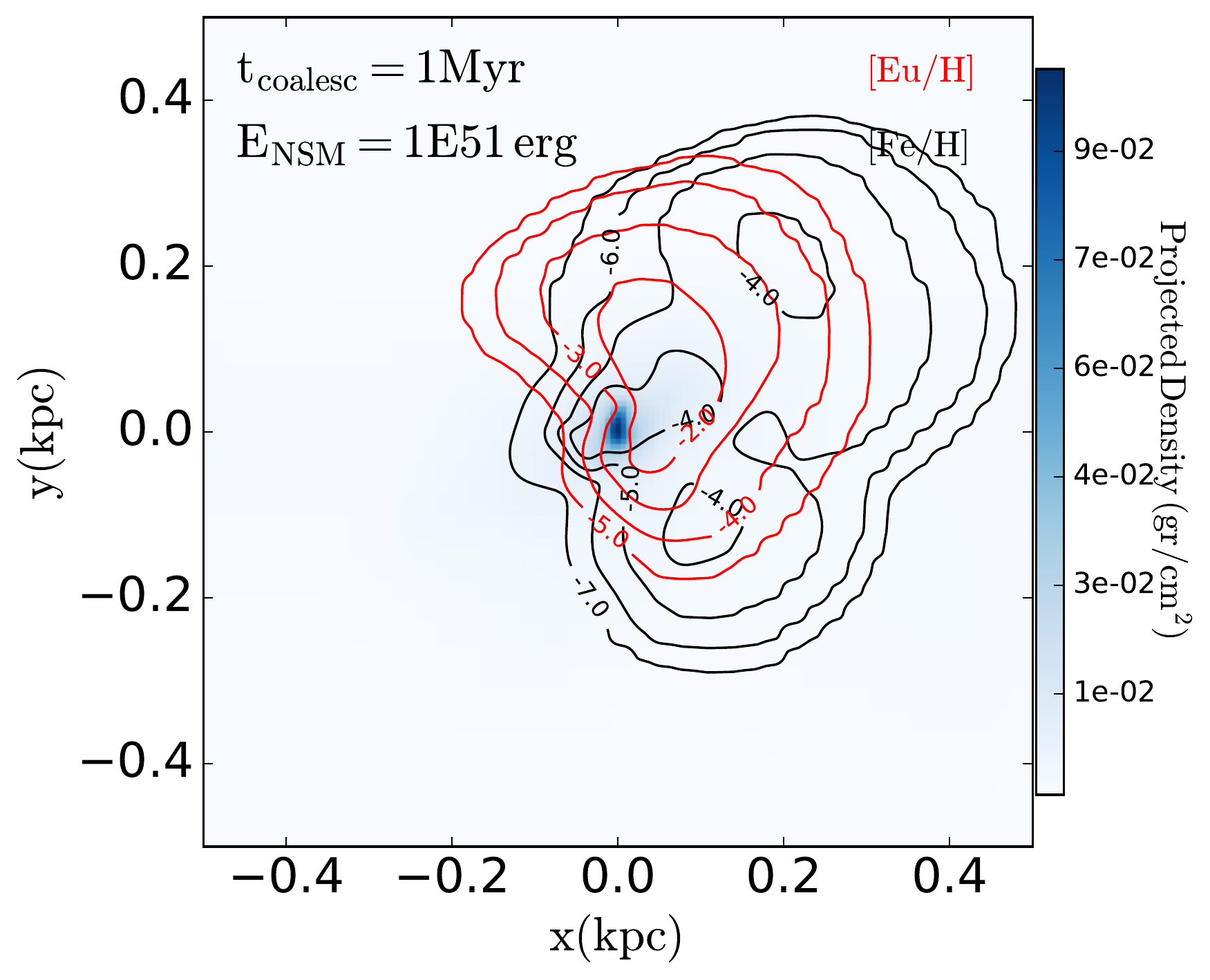}}
\resizebox{2.1in}{!}{\includegraphics[]{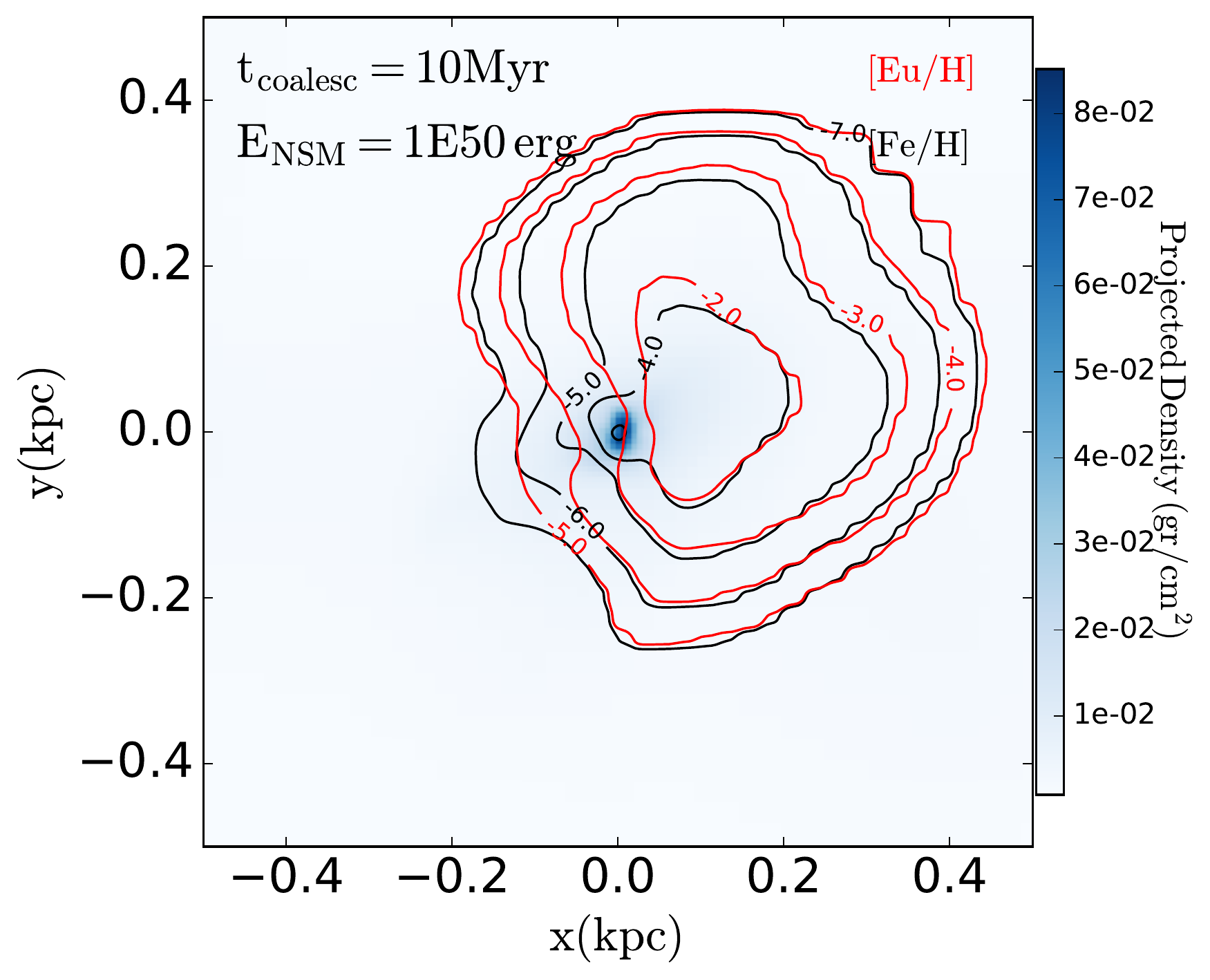}}
\resizebox{2.1in}{!}{\includegraphics[]{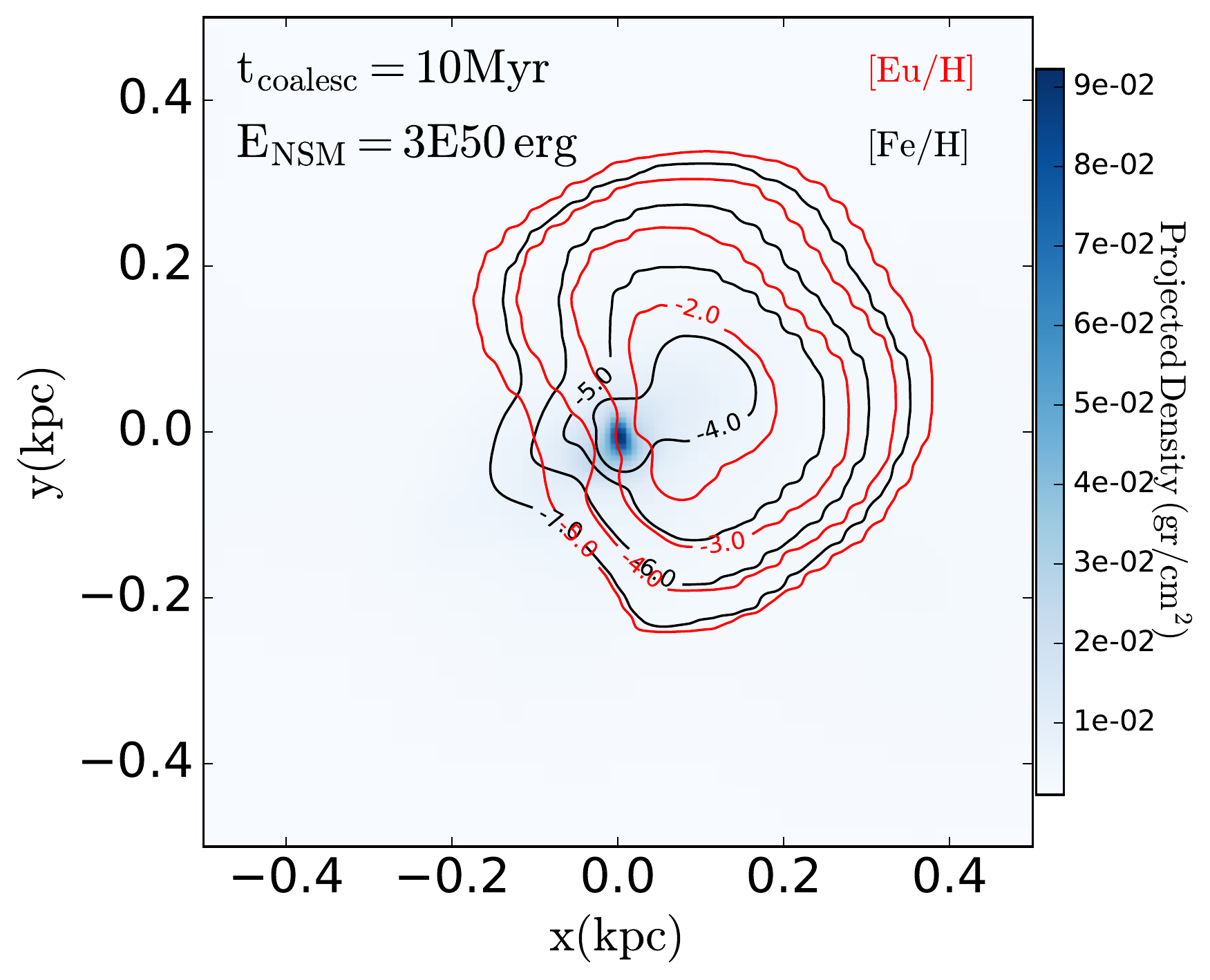}}
\resizebox{2.1in}{!}{\includegraphics[]{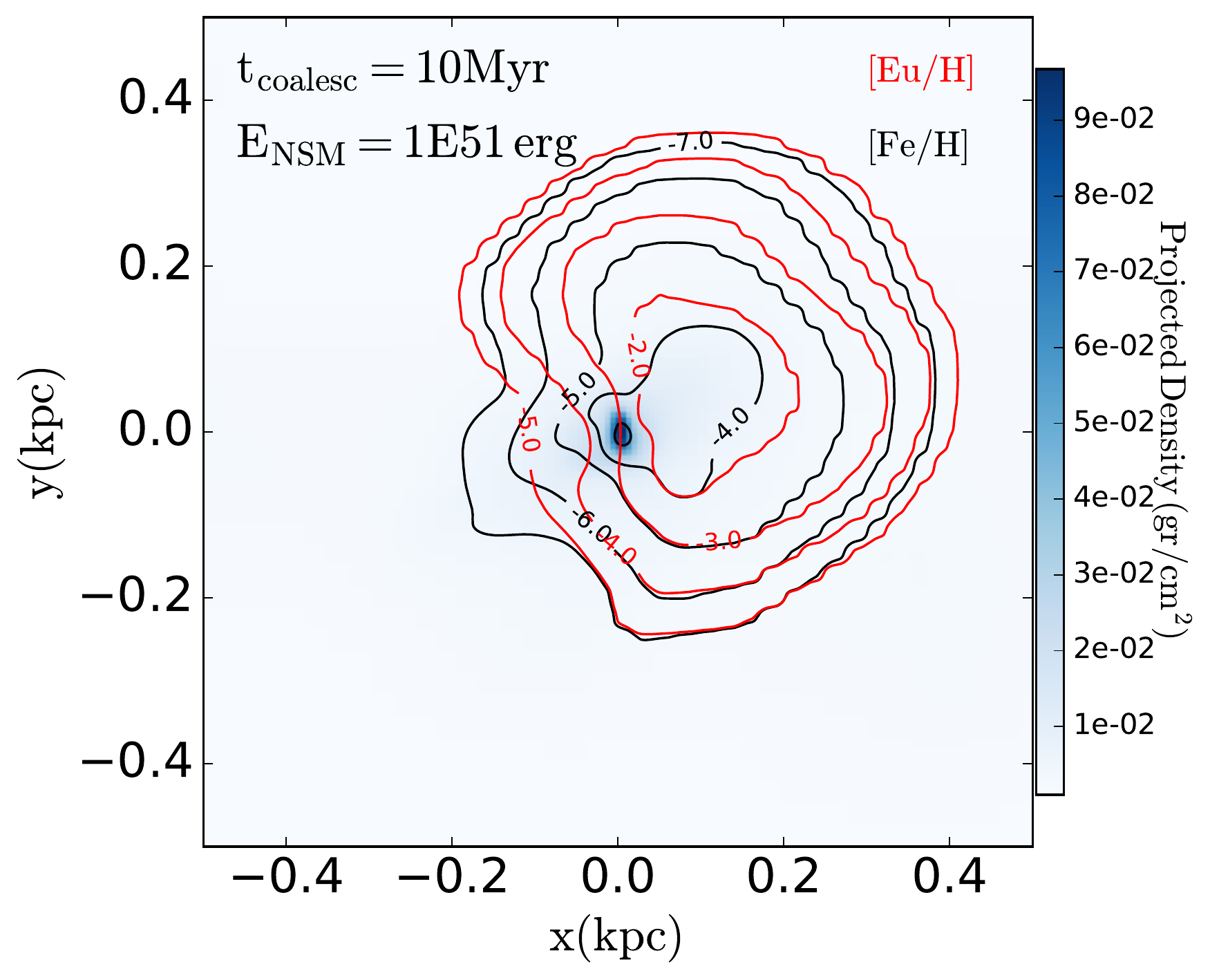}}
\caption{The projected density of our second UFD candidate at $z=12.3$. As in Figure 1, contours of the gas phase abundances of [Fe/H] (black) and [Eu/H] (red) are overlaid,  and from left to right and from top to bottom  \ensm = $10^{50}$, $3 \times 10^{50},$ and  $10^{51}$\,erg and \tcoalesc = 1 and 10\,Myr respectively.  Also as in Figure 1, $M_r=10^{-3} M_{\odot}$ for all cases. In this galaxy, the off-center NSM event leads to a larger spatial distribution of ejecta, which makes the polluted gas more diluted. This leads to lower r-process enrichment of the next generation of the stars than seen in the case in which  r-process material is dispersed more locally. When \tcoalesc =1 Myr, the NSM occurs at $z=13.24.$
The NSM is a specific particle 
that is tagged in our simulation and therefore unique for a given set of simulation, however the assignment of SNcc to stellar particles is done in a stochastic fashion
and therefore the contours of [Fe/H] can be different for each simulation. }
\vspace{-0.3cm}
\label{high_z_gas}
\end{figure*}

\begin{figure*}
\centering
\resizebox{3.2in}{!}{\includegraphics[]{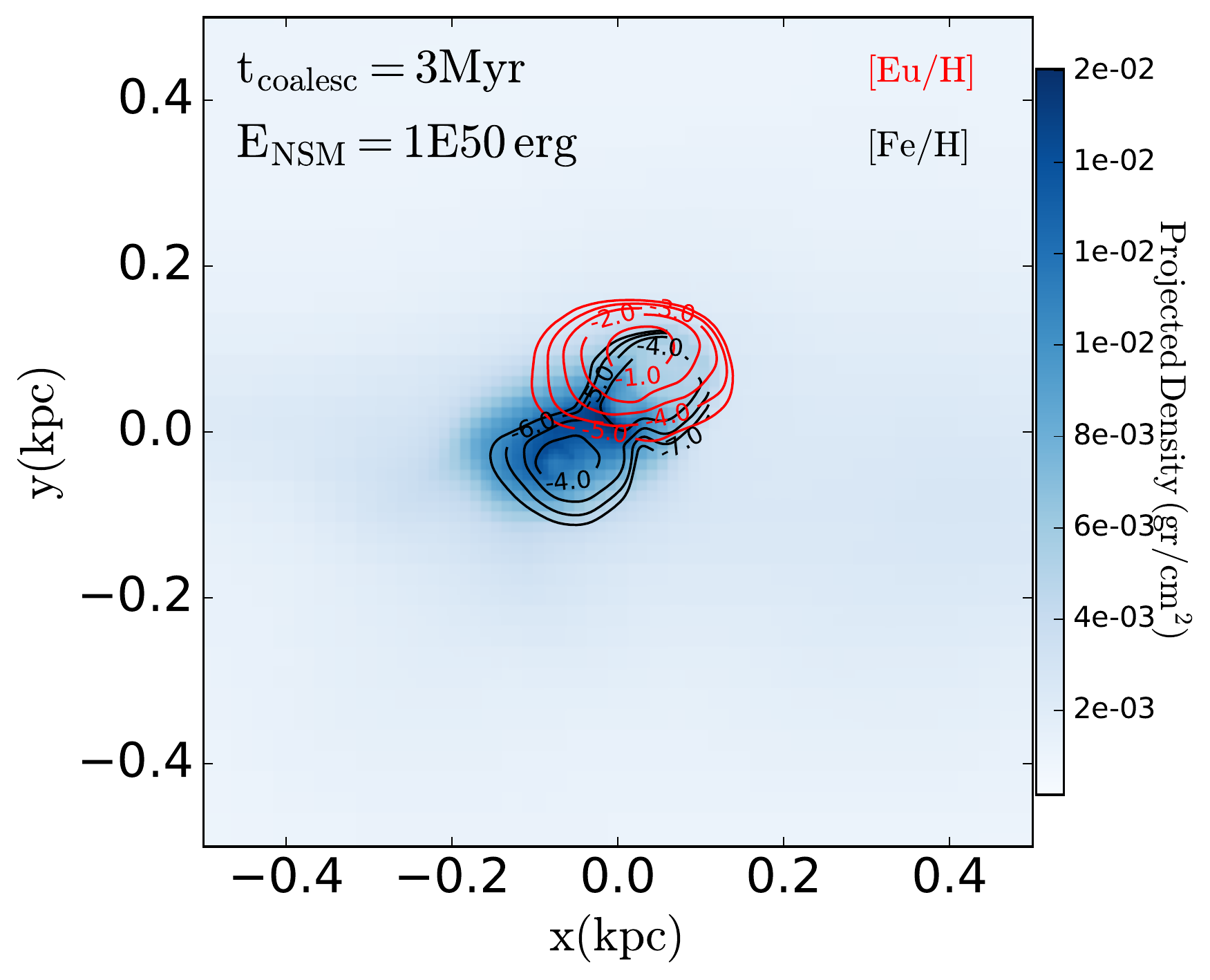}}
\resizebox{3.2in}{!}{\includegraphics[]{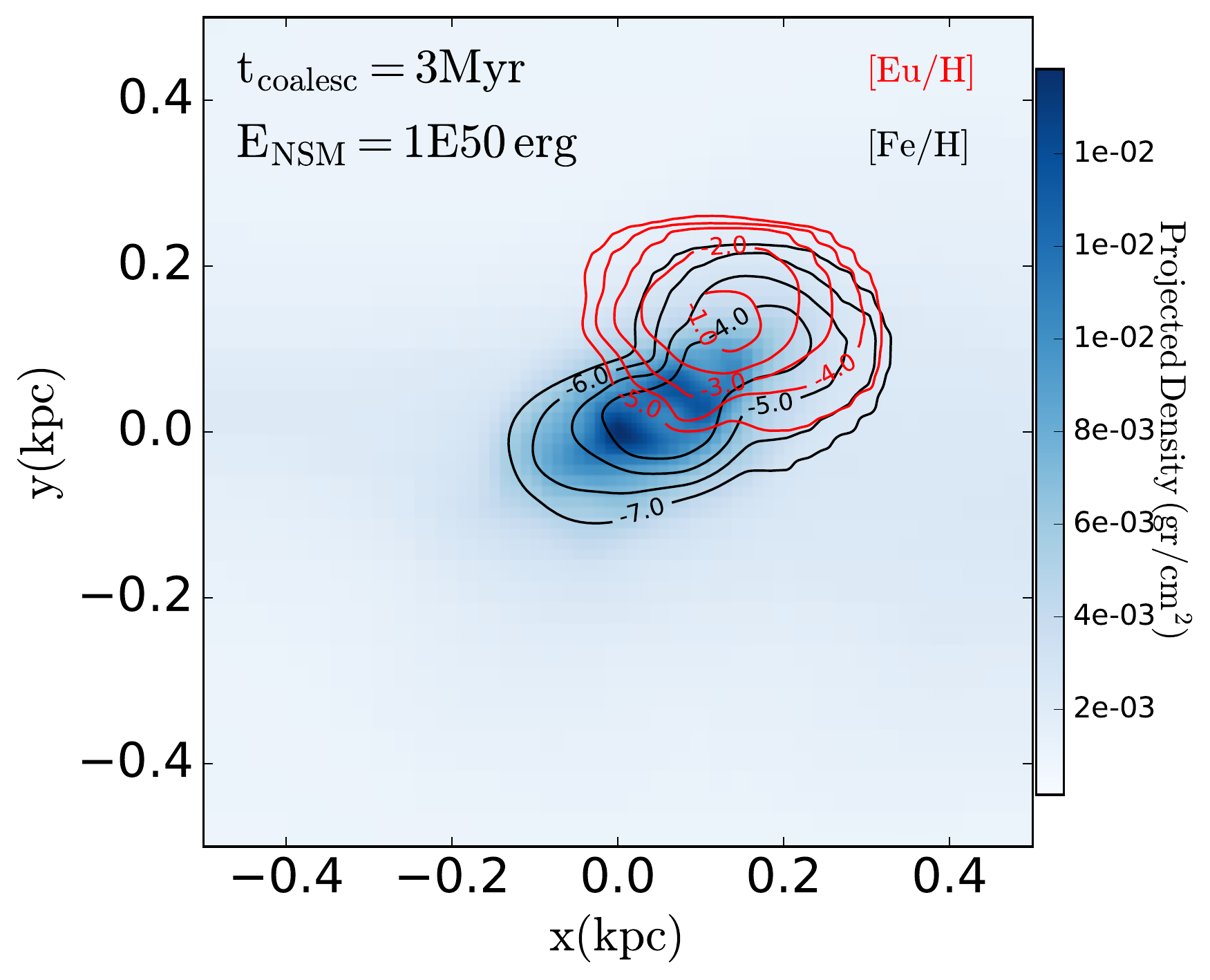}}
\caption{The same as Figure 2 but showing the results at earlier redshifts:  ($z=12.9,$ left) and $z=12.7$ (right).
The NSM event happens at $z=13.16$ in a clump of gas which is less dense than the larger clump into which it is merging.
 Such off-center NSM events lead to the expansion of the r-process material over a larger volume as compared with events that occur 
in the densest part of the galaxy. Note the expansion of the gas compared to what we observed for our lower redshift system, is more in the direction of low density region (upper-right) while the lower left extent is similar to the result for the lower redshift system. }
\vspace{-0.3cm}
\label{high_z_evolution}
\end{figure*}

\begin{figure*}
\centering
\resizebox{2.1in}{!}{\includegraphics[]{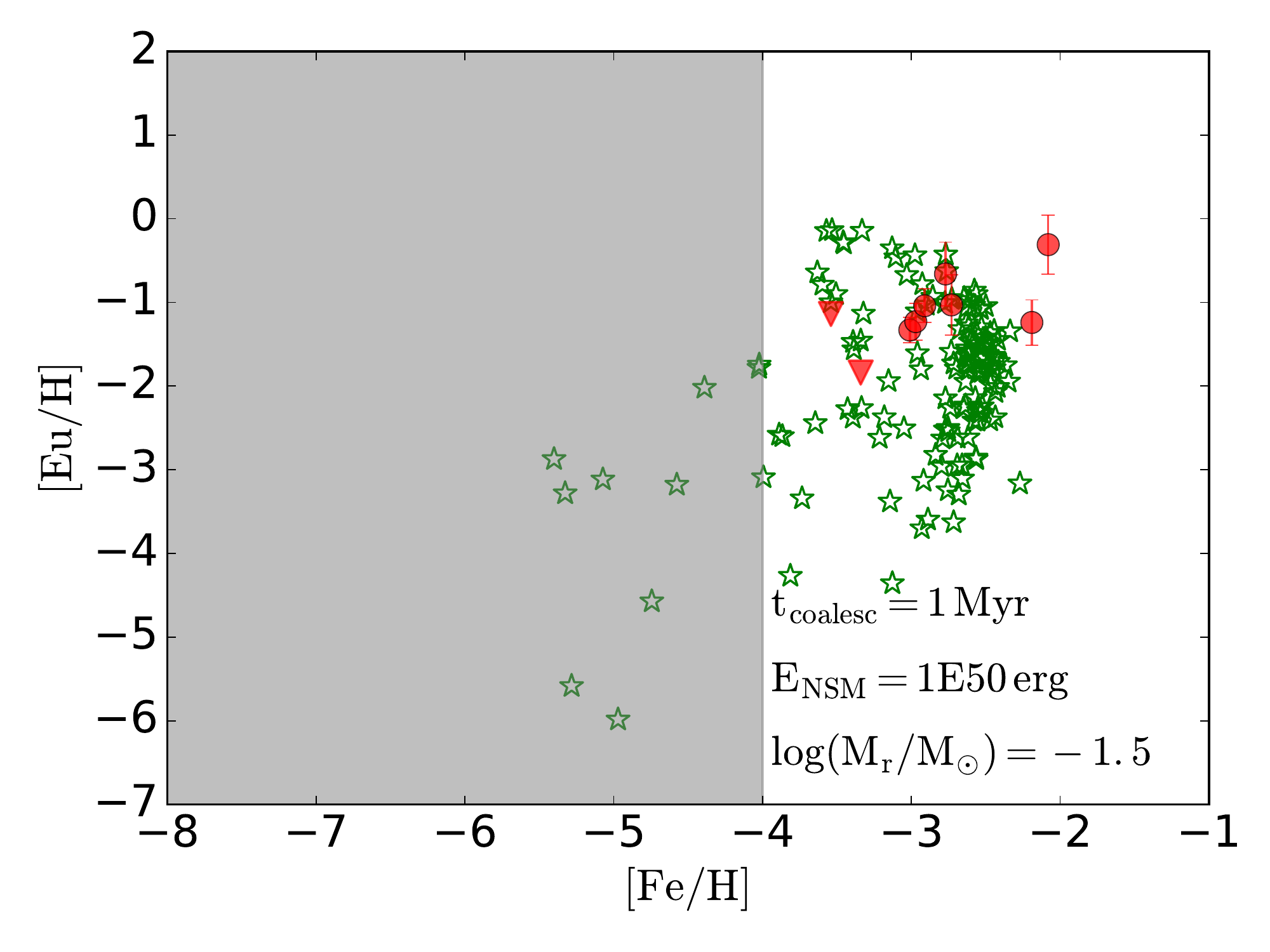}}
\resizebox{2.1in}{!}{\includegraphics[]{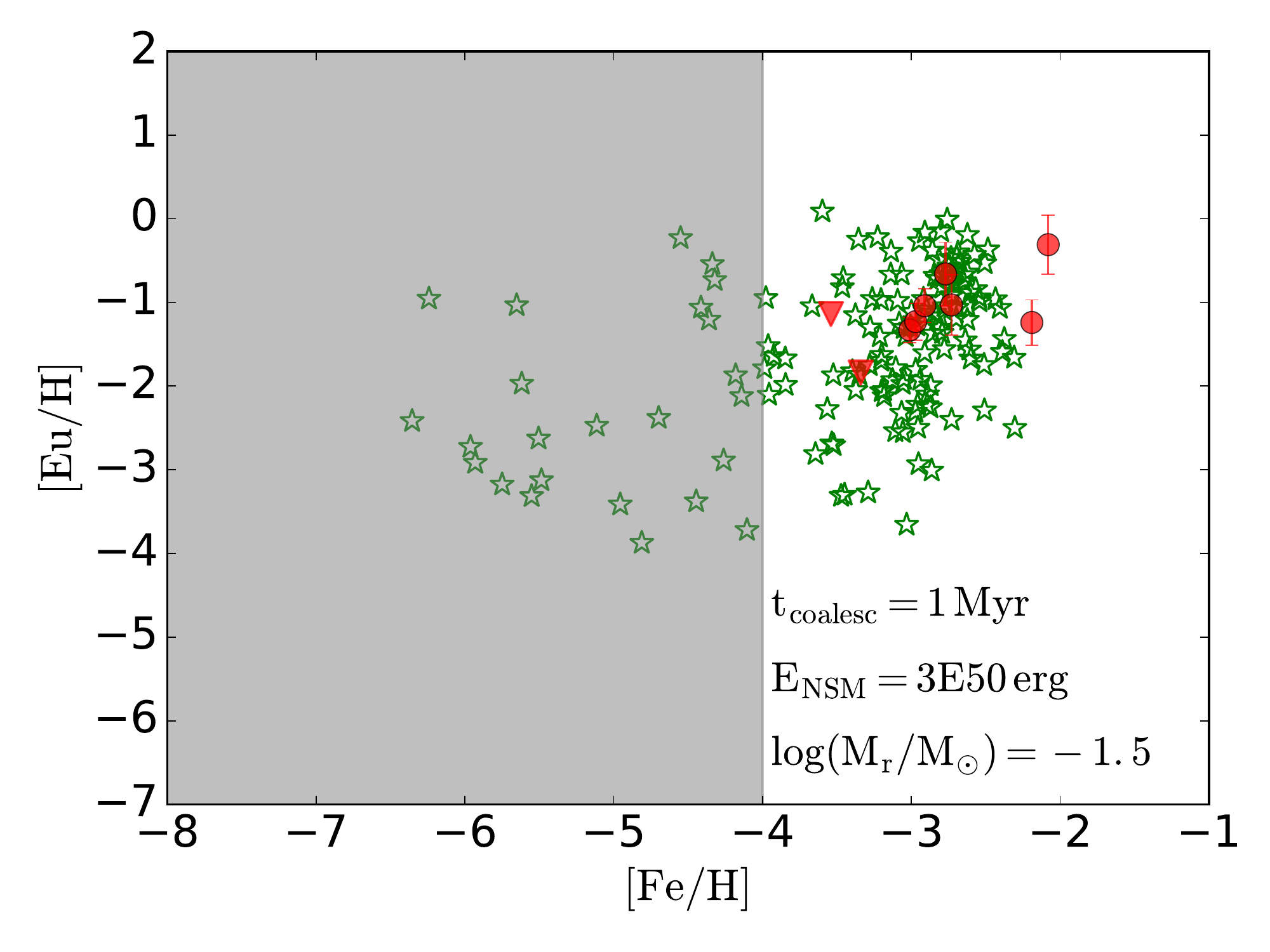}}
\resizebox{2.1in}{!}{\includegraphics[]{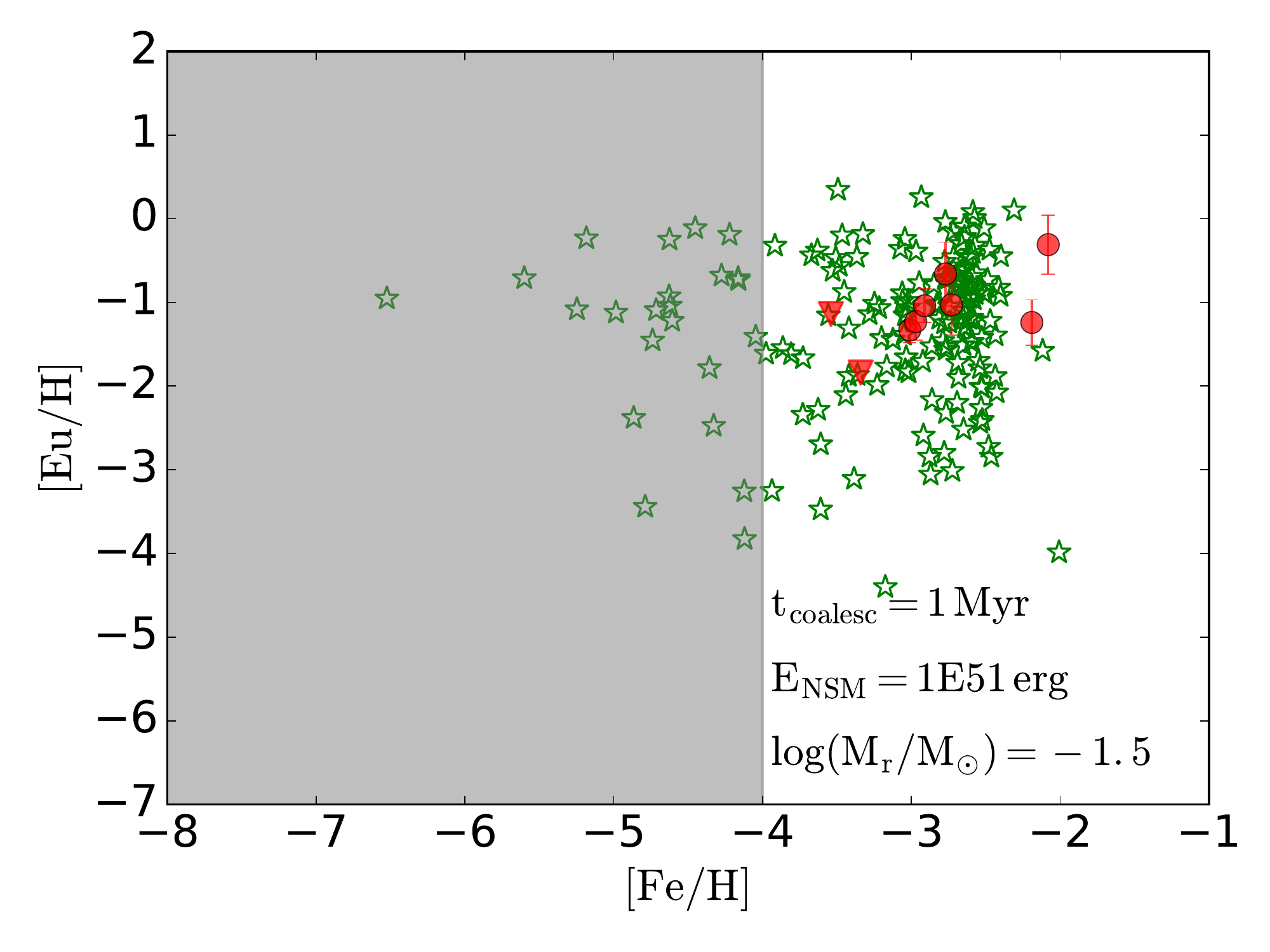}}
\resizebox{2.1in}{!}{\includegraphics[]{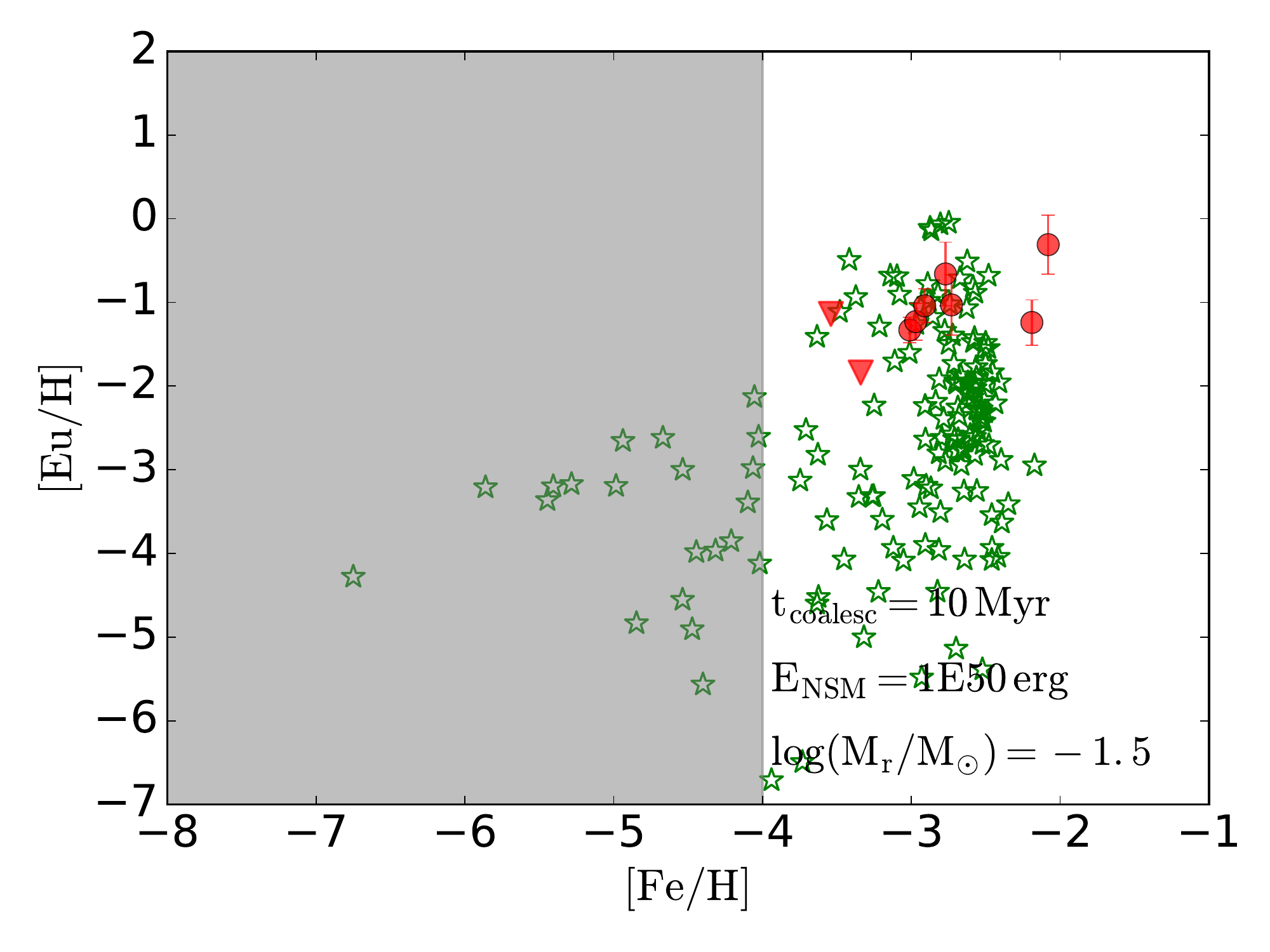}}
\resizebox{2.1in}{!}{\includegraphics[]{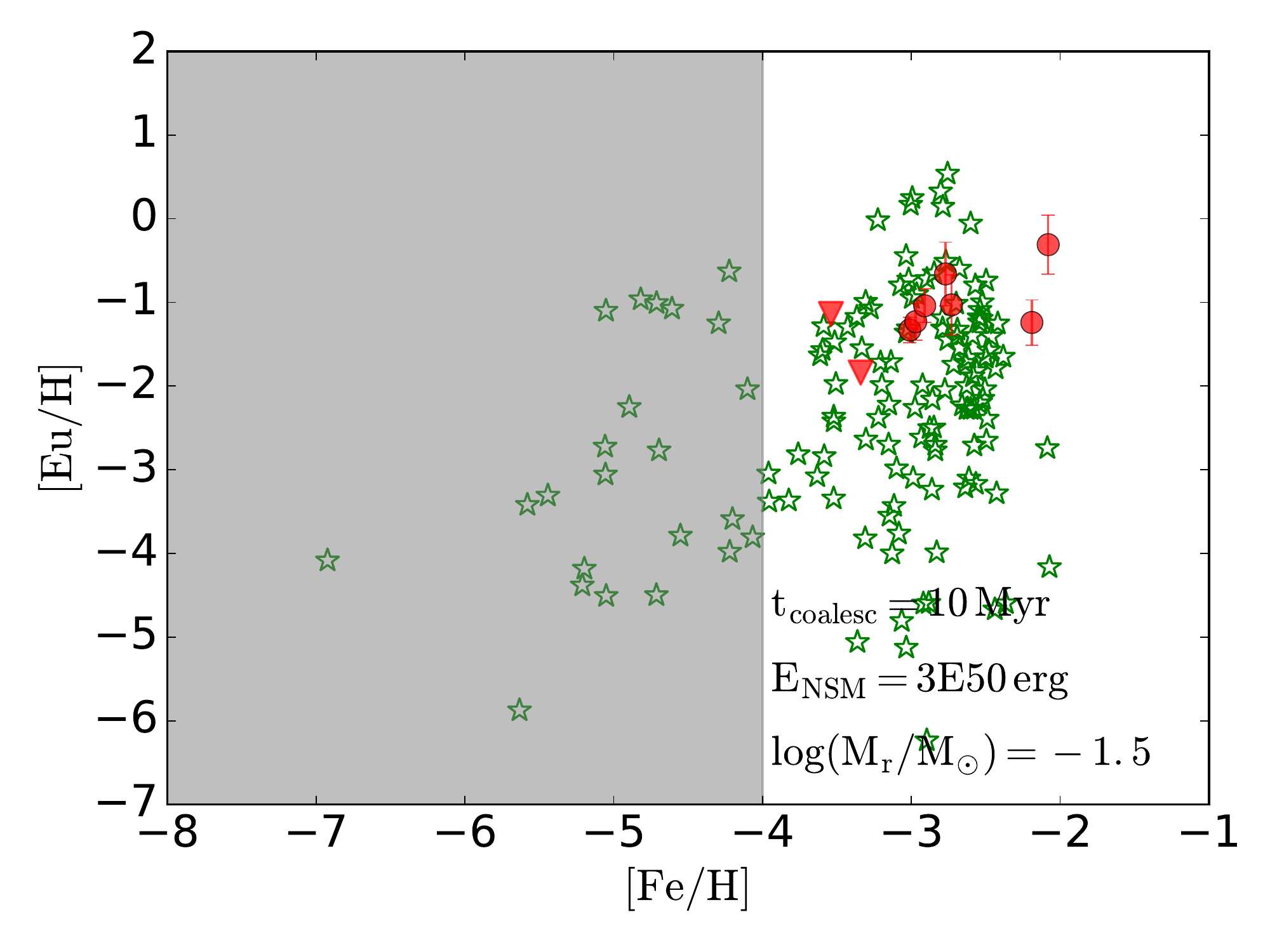}}
\resizebox{2.1in}{!}{\includegraphics[]{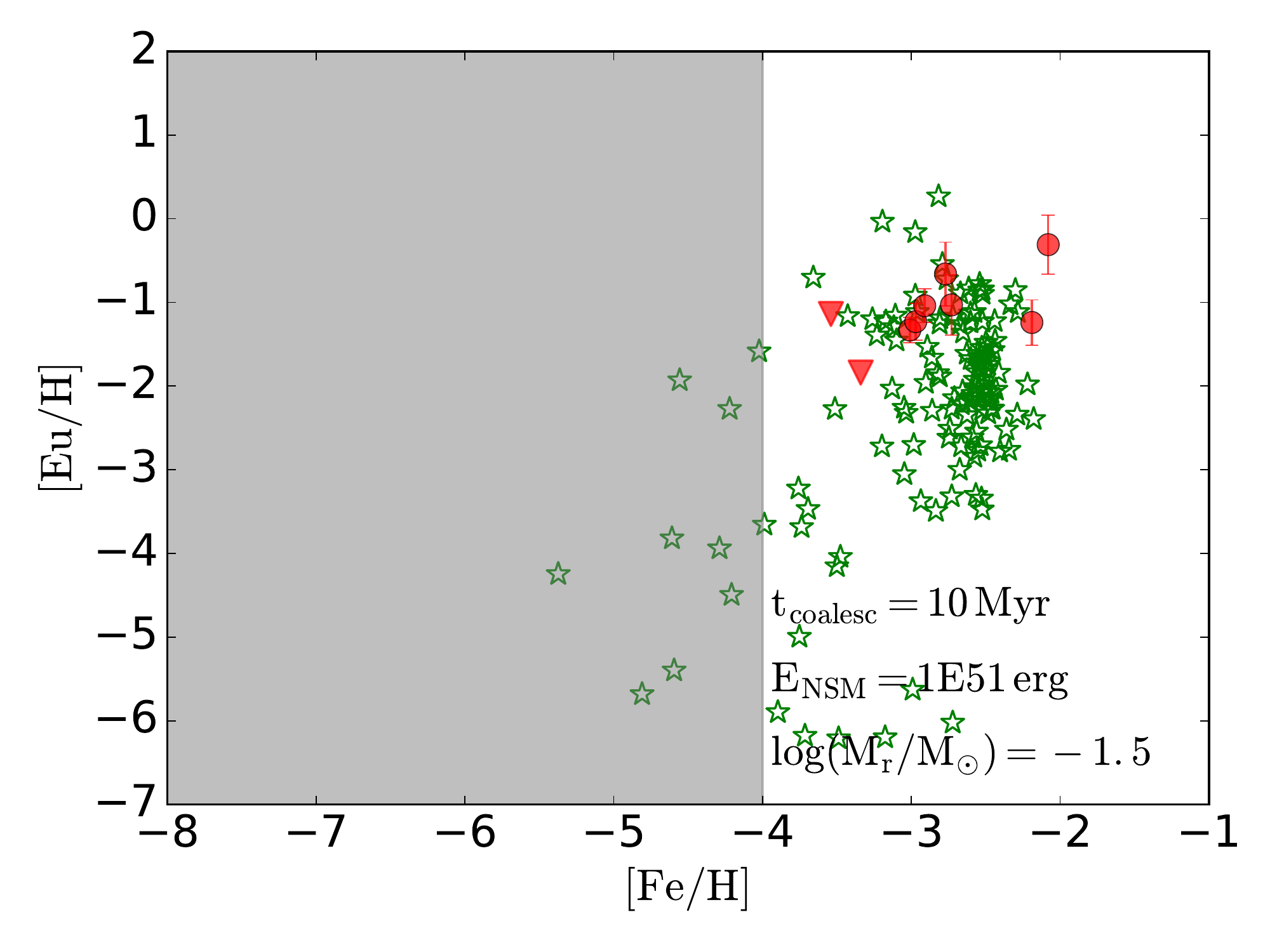}}
\caption{The same as Figure 2 but showing the results for the other system we simulated at $z=12.3$. 
n all cases, the r-process material mass ejected ($M_r$) in a NSM is mentioned in the panel.
The stellar particles are not enough enriched in Eu compared with the RetII stars. 
Here we show the distribution when the best KS test result is achieved with ($\log (M_r/M_{\odot})=-1.5$) which is the maximum allowed by theoretical studies.
However, acceptable agreement with the data based on 2D KS test is never achieved even when we adopt the highest $M_r$ possible. 
The solid red points are the nine stars in Ret II system which is a local UDF heavily enriched by r-process elements \citep{Ji:2016ja}. Two most metal-poor stars in Ret II are upper limits for [Eu/H], not detections.
}
\vspace{-0.3cm}
\label{high_z_gas}
\end{figure*}

In order to capture the minimum variance of our results, we performed the same suite of zoom simulations on another halo, in  which  star formation begins at higher redshifts. 
Figure 4 shows the distribution of gas, metals, and r-process elements in our higher-redshift galaxy at $z=12.3.$ 
Note that the spatial extent of the iron and r-process elements is larger than in the galaxy presented in Figure 1. 
This is due to the peculiar star formation history of this galaxy, which begins as two merging clumps of gas that start forming stars within a few Myrs of each other.  The NSM occurs in the less dense clump, and this causes the  r-process elements to be dispersed over a much larger volume than if it had gone off in the dense (central) part of the potential, as was the case in the $z\approx 8$ galaxy. The unusual star formation history of this halo is illustrated in Figure 5, which shows the same halo at earlier times, when the r-process has just happened in the gas clump displaced from the densest part of the galaxy.

This leads the stellar particles to be insufficiently enriched in Eu compared to the nine stars of the Ret II as shown in Figure 6.
By performing a 2D KS test we find that we would need to inject about two orders of magnitude more r-process mass into the ISM for the NSM event to reach P values of $\approx 0.1$ against the RetII data points. 
However, $M_r=10^{-1}M_{\odot}$ is beyond the plausible range for the r-process mass ejecta as is suggested by the simulations. The P value with $M_r=4\times10^{-2} M_{\odot}$ ranges from $\approx 10^{-3}$ to $10^{-4}$ for this system among
different models. 

Although the off-center NSM event occurs randomly in our simulation, such events could happen as a result of large natal kicks of the neutron star merger \citep{Beniamini:2016kw,Bramante:2016kp}. For example in our case the spatial off-set is about 0.1 kpc, which could arise due to kicks with velocities in the range of $\rm 10-100 km/s$ assuming a typical timescale of 1-10 Myr for the system. Such high natal kicks put the NSM event well outside the dense parts of the galaxy and therefore lead to low r-process enrichment of the subsequent stars formed in the system, a picture that would not be able to explain the Ret II observations.

The impact of \tcoalesc on the [Eu/H]-[Fe/H] plane for three different sets of star formation histories has been studied by \citet{Vangioni:2016jj}. They find that at a fixed [Fe/H], higher coalescence  timescales lead to lower values of [Eu/H]. We do not observe such trend in our simulation, in that our lower redshift system shows relatively higher values of [Eu/H] at a fixed [Fe/H] as the \tcoalesc is increased.  Moreover, there is no trend for our higher redshift system with \tcoalesc but we know the off-center NSM event has had a large impact in that systems' stellar particles' abundance in [Eu/H]-[Fe/H] plane. However, the coalescence timescale studied by \citet{Vangioni:2016jj} is 0,0.05,0.1 and 0.2 Gyr where only the \tcoalesc=0 Myr case could be compared to our results. The coalescence timescale range that we have studied may not cover a wide enough range so we can observe its impact in our results. However, we can not study longer timescales because otherwise the NSM would occur after the star formation of the system has ceased and therefore
we would not see any r-process enrichment in the stars.

\section{Summary and Conclusions}

Highly-enriched local UDFs like Ret II are perfect candidates for studying the production sites  of r-process elements. Star formation in these systems is quenched by reionization  such that  r-process enrichment must have occurred at $z>z_{\rm reion}$.  Neutron star mergers have long been a promising candidate for the r-process element production, and for systems such as RetII, only one such event could explain the stellar abundances observed.

We performed cosmological zoom simulations of two different halos, both with a mass of $M_H\approx10^8 M_{\odot}$ at $z=6$. Each included one NSM at the very early stages of its star formation history, modeled as a stellar particle that hosts two SNcc, creating a neutron star binary that merges at a time \tcoalesc time later.  We chose a star particle mass $\approx 50 M_{\odot}$ to both resolve the stellar content of the RetII like system ($\approx10^4 M_{\odot}$) and make it possible for a single particle to host the two SNcc events required to produce a NSM. Since the stellar particle mass in our simulation was small, we modeled supernova feedback stochastically for all the other stellar particles in the system, with a SNcc occuring  50\% of the time. We modeled the NSM event with two variables: the energy of the NSM event ($E_{\rm NSM}$), which we varied between $10^{50}-10^{51}$ \erg, and the coalescence timescale (\tcoalesc) that we varied between 1 and 10 Myr. When post-processing our simulations, we allowed the amount of r-process material released into the ISM in a NSM event to vary between $10^{-4}M_{\odot}$ and $4\times10^{-2}M_{\odot},$ with a fiducial values of  $10^{-3}M_{\odot},$ and we converted the r-process content to europium, the primary element that is almost solely produced in r-process events.  The Eu yield adopted in this study is comparable to what \citet{Vangioni:2016jj} had implemented ($7\times10^{-5}M_{\odot}$) by considering the yields from both the binary merger phase and the BH-torus evolutions. 

Our results show that a single NSM  can lead to a distribution of stellar r-process abundances similar to those observed in Ret II.  In one of the two halos that we simulated, the NSM event took place at the center of the stellar distribution, leading to a high spatial correlation between the r-process material and the supernovae ejecta.  This not only lead to high-levels of r-process enrichment  such as seen in Ret II, but also the positive correlation between [Eu/H] and [Fe/H].   In the second halo, the NSM event took place away from the densest part of the galaxy, and the r-process material expanded primarily into the low density ISM.  In this case, the more extended and shallower r-process distribution lead to stars that were  on average about 2 dex under-abundant in europium as compared with Ret II stars. Although even in this case we still see some of the stars with high levels enrichment comparable to those of Ret II, our simulations show that even without modeling the natal kicks, the binary can explode in places that would lead to very inefficient r-process enrichment in the system.

Thus it is the location of NSM event, rather than the ejection energy or the  coalescence delay time scale, that is the dominant parameter in determining the r-process distribution in ultra faint dwarf galaxies.  This means that hydrodynamic simulations, such as the one carried out in this study are required to reliably interpret measurements of r-process enhanced metal poor stars.  It also means that the natal kicks of neutron stars, which were not modeled in this study, are likely to play in important role in determining these abundances.  Further theoretical modeling of this processes is needed to better understand the r-process enrichment of local ultra-faint dwarf  galaxies.

Recently, \citet{Hansen:2017ho} reported a discovery of a star with [Eu/H]=-1.65 and [Fe/H]=-2.25 in Tucana III. The stellar particles in our simulation cover a wide range in [Eu/H] -[Fe/H] plane and the star that \citet{Hansen:2017ho} observed lies where we have simulated stars to match its abundance. Both our high and low redshift simulated UFD's stars easily overlap with this star, therefore we can not exclude the possibility that a NSM merger is also responsible for its enrichment. However more than one star is required to have a sufficient statistics to robustly constrain the origin of r-process enhancement in Tucana III.

Objects with $\vpeak \leq 25 \kms$ and $\Mpeak > 10^{7.5}M_{\odot}$ are potential candidates to survive as present day UFDs. All ``peak'' quantities are defined as occurring at the time at which a halo's main branch reaches its maximum mass; i.e. $\Mpeak$ is the maximum mass of a halo, and $\vpeak$ is $\vmax$ at that time. This criterion is set to define these systems as true fossils in that they never reached a mass scale in their entire history large enough to be able to accrete gas from IGM and therefore reignite star formation and be considered as polluted systems \citep{Gnedin:2006gl,Bovill:2011bk}. 
To see whether our simulated dwarfs will survive as present day UFDs, it would be necessary to follow their evolution down to $z=0$. Moreover the candidates would need to be  selected from a MW progenitor. In a parallel paper (Safarzadeh \& Ji, in prep) we will present the statistics of such systems on the probability of a random selection of a $10^8 M_{\odot}$ halo at z=6 to survive as a present day UFD. The probabilities are about $\sim 10\%$ consistent with \citet{Gnedin:2006gl}.

Short gamma ray bursts (sGRBs) are believed to be the result of neutron star mergers and the coalescence timescales considered in this study are on the short end of the allowed range based on sGRB observations \citep{Berger:2010de}
and population synthesis models. The coalescence time can be very long,  \citep[e.g.][]{Dominik:2012cw} and the delay time distributions could be modeled  as a power law ($\propto t^{-1}$) with  minimum timescale of 1 Myr and maximum of 10 Gyr. Clearly if the NSM occurs after the star formation history of the system has ceased due to re-ionization, then there would be no sign of r-process enrichment in the stars of such a UFD. It can be that NSM has occurred in other UFDs but at a time at which there is no subsequent star formation. We have shown that the dominant variable is the location of the NSM in the system, which determines the subsequent enrichment of the newly born stars. Moreover, not only the coalescence timescale, but also the natal kick distribution would impact the possibility of r-process enrichment in a system. This combined effect will be explored in future work.

\section{Acknowledgements}
We are thankful to the anonymous referee for their careful reading of our manuscript. We are thankful to Frank Timmes, Alexander Ji, Mark Richardson and Rick Sarmento for useful discussions. 
We used the yt package \citep{Turk:2011dd} and pynbody \citep{Pontzen:2013tn} for part of the analysis in this work.
This work was supported by the National Science Foundation under grant AST14-07835 and by NASA under theory grant NNX15AK82G. 
We would also like to thank the Texas Advanced Computing Center (TACC) at The University of Texas at Austin, and the Extreme Science and Engineering Discovery Environment (XSEDE) for providing HPC resources via grant TG-AST130021 and TG-AST160063 that have contributed to the results reported within this paper.

\bibliographystyle{mnras}
\bibliography{the_entire_lib.bib}

\bsp
\label{lastpage}

\end{document}